\documentclass[a4paper,12pt]{article}

\usepackage{helvet,url}

\usepackage{amsthm}
\usepackage{graphicx, amsmath, amssymb}
\usepackage{amsfonts, enumerate}

\usepackage{multirow}

\newcommand{\ignore}[1]{}

\newcommand{\no}[2]{#1}

\newcommand{\xml}{{\tt xml}}
\newcommand{\dna}{{\tt dna}}
\newcommand{\english}{{\tt english}}
\newcommand{\pitches}{{\tt pitches}}
\newcommand{\proteins}{{\tt proteins}}
\newcommand{\sources}{{\tt sources}}

\newcommand{\lzindex}{{\tt LZ-index}}
\newcommand{\sau}{{\tt plain SA}}
\newcommand{\csa}{{\tt CSA}}
\newcommand{\afindex}{{\tt AF-index}}
\newcommand{\ssa}{{\tt SSA}}
\newcommand{\FMInew}{{\sf FMI2}}

   {\vspace*{-4pt}
    \begin{enumerate}\itemsep=0pt}%
   {\end{enumerate}
    \vspace*{-2pt}}

   {\vspace*{-4pt}
    \begin{itemize}\itemsep=0pt}%
   {\end{itemize}
    \vspace*{-2pt}}

\newcommand{\hk}[1]{H_{#1}}     

\newcommand{\gzip}{{\sf gzip}}
\newcommand{\bzip}{{\sf bzip2}}
\newcommand{\ppm}{{\sf PPMDi}}

\DeclareTextCommand{\textbraceleft}{OT1}{\texttt{\symbol{`\{}}}
\DeclareTextCommand{\textbraceright}{OT1}{\texttt{\symbol{`\}}}}

\title{Compressed Text Indexes:From Theory to Practice!\thanks{Partially supported by Yahoo! Research grants "Data compression and 
indexing in hierarchical memories" (Barcelona) and "Compressed data 
structures" (Chile), Italian MIUR grants Italy-Israel FIRB "Pattern Discovery 
Algorithms in Discrete Structures, with Applications to Bioinformatics", PRIN 
"MainStream MAssive INformation structures and dataSTREAMs", and Millennium
Nucleus Center for Web Research, Grant P04-067-F, Mideplan, Chile.}}

\author{%
Paolo Ferragina $^1$ ~
Rodrigo Gonz\'alez $^2$ ~\\
Gonzalo Navarro $^2$ ~
Rossano Venturini $^1$  
\ \\
\ \\
\small $^1$ Dept. of Computer Science, University of Pisa.\\
{\tt \textbraceleft ferragina,rventurini\}@di.unipi.it} \\
\small
$^2$ Dept. of Computer Science, University of Chile.\\
{\tt \textbraceleft rgonzale,gnavarro\}@dcc.uchile.cl}}

\begin{document}

\maketitle
\begin{abstract}

A \textit{compressed full-text self-index} represents a text in a
compressed form and still answers queries efficiently. This
technology represents a breakthrough over the text indexing
techniques of the previous decade, whose indexes required several
times the size of the text. Although it is relatively new, this
technology has matured up to a point where theoretical research is
giving way to practical developments. Nonetheless this requires 
significant programming skills, a deep engineering effort, and a
strong algorithmic background to dig into the research results.
To date only isolated implementations and focused comparisons of 
compressed indexes have been reported, and they missed a
common API, which prevented their re-use or deployment within other 
applications.

The goal of this paper is to fill this gap. First, we present the existing
implementations of compressed indexes from a practitioner's point of view. 
Second, we introduce the {\em Pizza\&Chili} site, which offers tuned 
implementations and a standardized API for the most successful compressed 
full-text self-indexes, together with effective testbeds and scripts for 
their automatic validation and test. Third, we show the results of our
extensive experiments on these codes with the aim of demonstrating the 
practical relevance of this novel and exciting technology.
\end{abstract}

\section{Introduction}
A large fraction of the data we process every day consists of a
sequence of symbols over an alphabet, and hence is a {\em text}.
Unformatted natural language, XML and HTML collections, program
codes, music sequences, DNA and protein sequences, are just the
typical examples that come to our mind when thinking of text 
incarnations. Most of the manipulations required over those
sequences involve, sooner or later, {\em searching} those (usually 
long) sequences for (usually short) pattern sequences. Not
surprisingly, text searching and processing has been a central
issue in Computer Science research since its beginnings.

Despite the increase in processing speeds, sequential text
searching long ago ceased to be a viable alternative for many 
applications, and indexed text searching has become mandatory. A
{\em text index} is a data structure built over a text which
significantly speeds up searches for arbitrary patterns, at the
cost of some additional space. The inverted list structure
\cite{Inverted_list} is an extremely popular index to handle
so-called ``natural language'' text, due to its simplicity, low
space requirements, and fast query times. An inverted list is
essentially a table recording the occurrences of every distinct
text word. Thus every word query is already precomputed, and
phrase queries are carried out essentially via list intersections.
Although inverted lists are ubiquitous
in current Web search engines and IR systems, they present three
main limitations: (1) there must exist a clear concept of ``word''
in the text, easy to recognize automatically; (2) those words must
follow some statistical rules, so that there are not too many
different words in a text, otherwise the table is too large; (3)
one can search only for whole words or phrases, not for any
substring.

There are applications in which these limitations are
unacceptable --- e.g. bio-informatics, computational linguistics,
multimedia databases, search engines for agglutinating and Far
East languages --- and thus \textit{full-text indexing} must be
used. With this term we mean an index which is able to support so-called
{\em substring} searches, that is, searches not limited to word boundaries 
on any text $T$. These indexes will be the focus of our paper.

Although space consumption by itself is not really a problem given
the availability of cheap massive storage, the access speed of
that storage has not improved much, while CPU speeds have been
doubling every 18 months, as well the sizes of the various
(internal) memory levels. Given that nowadays an access to the
disk can be up to one million times slower than main memory, it is
often mandatory to fit the index in internal memory and leave as
few data as possible onto disk. Unfortunately, full-text indexing
was considered a technique that inevitably wasted a lot of space:
Data structures like suffix trees and suffix arrays required at
the very least four times the text size to achieve reasonable 
efficiency.

This situation is drastically changed in less than a decade
\cite{NMacmcs06}. Starting in the year 2000, a rapid sequence of 
achievements showed how to relate information theory with
string matching concepts, in a way that index regularities that
show up when the text is compressible were discovered and
exploited to reduce index occupancy without impairing the query
efficiency. The overall result has been the design of full-text
indexes whose size is proportional to that of the {\em compressed}
text. Moreover, those indexes are able to reproduce any text
portion without accessing the original text, and thus they {\em
replace} the text --- hence the name {\em self-indexes}. This way
{\em compressed full-text self-indexes} (compressed indexes, for
short) allow one to add search and random access functionalities to
compressed data with a negligible penalty in time and space
performance. For example, it is feasible today to index the 3 GB
Human genome on a 1 GB RAM desktop PC.

Although a comprehensive survey of these theoretical developments
has recently appeared \cite{NMacmcs06}, the algorithmic technology
underlying these compressed indexes requires for their
implementation a significant programming skill, a deep engineering
effort, and a strong algorithmic background. To date only
isolated implementations and focused comparisons of compressed
indexes have been reported, and they missed a common API,
which prevented their re-use or deploy within other applications.
The present paper has therefore a threefold purpose:


\medskip \noindent {\bf Algorithmic Engineering.}
We review the
most successful compressed indexes that have been implemented so
far, and present them in a way that may be useful for software
developers, by focusing on implementation choices as well on their
limitations. We think that this point of view complements
\cite{NMacmcs06} and fixes the state-of-the-art for this
technology, possibly stimulating improvements in the design of
such sophisticated algorithmic tools.

In addition, we introduce two novel implementations of compressed indexes.
These correspond to new versions of the FM-Index, one of which combines the
best existing theoretical guarantees with a competitive space/time tradeoff
in practice.

\medskip \noindent {\bf Experimental.}
We experimentally compare a selected
subset of implementations. This not only serves to help
programmers in choosing the best index for their needs, but also
gives a grasp of the practical relevance of this fascinating
technology.

\medskip \noindent {\bf Technology Transfer.}
We introduce the {\em Pizza\&Chili} site\footnote{Available at two
mirrors:  {\small \tt pizzachili.dcc.uchile.cl} and {\small \tt
pizzachili.di.unipi.it}}, which was developed with the aim
of providing publicly available implementations of compressed
indexes. Each implementation is well-tuned and adheres to a
suitable API of functions which should, in our intention, allow
any programmer to easily plug the provided compressed indexes
within his/her own software. The site also offers a collection of
texts and tools for experimenting and validating the proposed
compressed indexes. We hope that this simple API and the good
performance of those indexes will spread their use in several
applications.


The use of compressed indexes is obviously not limited to
plain text searching. Every time one needs to store a set of
strings which must be subsequently accessed for query-driven or
id-driven string retrieval, one can use a compressed index with
the goal of squeezing the dictionary space without slowing down
the query performance. This is the subtle need that any programmer
faces when implementing hash tables, tries or other indexing
data structures. Actually, the use of compressed indexes has been
successfully extended to handle several other more sophisticated
data structures, such as dictionary indexes \cite{FV07-SIGIR},
labeled trees \cite{FLMM05_labeled_trees,WWW06}, graphs
\cite{N07_graphs}, etc. Dealing with all those applications is
out of the scope of this paper, whose main goal is to address the
above three issues, and comment on the experimental
behavior of this new algorithmic technology.

This paper is organized as follows. Section~\ref{sec:basics}
explains the key conceptual ideas underlying the most relevant
compressed indexes. Section~\ref{sec:indexes} describes how the
indexes implement those basic ideas. Section~\ref{sec:site}
presents the {\em Pizza\&Chili} site, and next
Section~\ref{sec:exper} comments on a large suite of experiments
aimed at comparing the most successful implementations of the
compressed indexes present in this site. Finally,
Section~\ref{sec:concl} concludes and explores the future of the
area.

\section{Basic Concepts}
\label{sec:basics}

Let us introduce some notation. We will refer to {\em strings}
with $S=S[1,\ell]=S_{1,\ell}=s_1s_2\ldots s_\ell$ to denote a
sequence of symbols over an alphabet $\Sigma$ of size $\sigma$. By
$S[i,j]=S_{i,j}=s_i s_{i+1}\ldots s_j$ we will denote substrings
of $S$, which are called prefixes if $i=1$ or suffixes if
$j=\ell$. The length of a string will be written
$|S|=|S_{1,\ell}|=\ell$, and the reverse of a string will be
written $S^r = s_\ell s_{\ell-1} \ldots s_1$.

The {\em text searching problem} is then stated as follows. Given
a {\em text} string $T[1,n]$ and a {\em pattern} $P[1,m]$, we wish
to answer the following queries: (1) {\em count} the number of
occurrences ($occ$) of $P$ in $T$; (2) {\em locate} the $occ$ positions in 
$T$ where $P$ occurs. In this paper we assume that $T$ can be
preprocessed, and an {\em index} is built on it, in order to
speed up the execution of subsequent queries. We assume that the 
cost of index construction is amortized over sufficiently many
searches, as otherwise sequential searching is preferable.

In the case of self-indexes, which replace the text, a third
operation of interest is (3) {\em extract} the substring
$T_{l,r}$, given positions $l$ and $r$ in $T$.

For technical convenience we will assume that the last text
character is $t_n = \$$, a special end-marker symbol that belongs
to $\Sigma$ but does not appear elsewhere in $T$ nor $P$, and that
is lexicographically smaller than any other symbol in $\Sigma$.

\subsection{Classical Full-Text Indexes}
\label{sec:classic}

Many different indexing data structures have been proposed in the literature
for text searching, most notably suffix trees and suffix arrays.

The {\em suffix tree} \cite{Gus97} of a text $T$ is a trie (or
digital tree) built on all the $n$ suffixes $T_{i,n}$ of $T$,
where unary paths are compressed to ensure $O(n)$ size. The suffix
tree has $n$ leaves, each corresponding to a suffix of $T$, and
each internal suffix tree node corresponds to a unique substring
of $T$ that appears more than once. The suffix tree can count the
pattern occurrences in time $O(m)$, independent of $n$ and $occ$,
by descending in the tree according to the successive symbols of
$P$ (each node should store the number of leaves that descend from
it). Afterwards, it can locate the occurrences in optimal $O(occ)$
time by traversing the subtree of the node arrived at counting.
The suffix tree, however, uses much more space than the text
itself. In theoretical terms, it uses $\Theta(n\log n)$ bits
whereas the text needs $n \log\sigma$ bits (logarithms are in base
2 unless otherwise stated). In practice, a suffix tree requires
from 10 to 20 times the text size.

The {\em suffix array} \cite{Manber_Myers93} is a compact version
of the suffix tree. It still requires $\Theta(n\log n)$ bits, but
the constant is smaller: 4 times the text size in
practice. The suffix array $A[1,n]$ of a text $T_{1,n}$ contains
all the starting positions of the suffixes of $T$ listed in
lexicographical order, that is, $T_{A[1],n} < T_{A[2],n} <\ldots <
T_{A[n],n}$. $A$ can be obtained by traversing the leaves of the
suffix tree, or it can be built directly by naive or sophisticated
ad-hoc sorting methods \cite{PST06}.

Any substring of $T$ is the prefix of a text suffix, thus finding
all the occurrences of $P$ is equivalent to finding all the text
suffixes that start with $P$. Those form a lexicographical
interval in $A$, which can be binary searched in $O(m\log n)$
time, as each comparison in the binary search requires examining
up to $m$ symbols of the pattern and a text suffix. The time can
be boosted to $O(m+\log n)$, by using an auxiliary structure that
doubles the space requirement of the suffix array
\cite{Manber_Myers93}, or even to $O(m+\log |\Sigma|)$ by adding
some further data structures (called {\em suffix trays}
\cite{Coleetal-icalp06}). Once the interval $A[sp,ep]$ containing
all the text suffixes starting with $P$ has been identified,
counting is solved as $occ = ep-sp+1$, and the occurrences are
located at $A[sp],~ A[sp+1], \ldots A[ep]$.

\subsection{Backward Search}
\label{sec:bsearch}

In the previous section we described the classical binary-search
method over suffix arrays. Here we review an alternative approach
which has been recently proposed in \cite{FM05}, hereafter named
{\em backward search}. For any $i=m, m-1, \ldots, 1$, this search
algorithm keeps the interval $A[sp_i,ep_i]$ storing all text
suffixes which are prefixed by $P_{i,m}$. This is done via two
main steps:

{\bf Initial step.}
We have $i=m$, so that it suffices
to access a precomputed table that stores the pair $\langle sp_m,
ep_m\rangle$ for all possible symbols $p_m \in \Sigma$.

{\bf Inductive step.} 
Let us assume to have computed
the interval $A[sp_{i+1},ep_{i+1}]$, whose suffixes are prefixed
by $P_{i+1,m}$. The present step determines the next interval
$A[sp_i,ep_i]$ for $P_{i,m}$ from the previous interval
and the next pattern symbol $p_{i}$. The implementation is not
obvious, and leads to different realizations of backward searching
in several compressed indexes, with various time performances.


The backward-search algorithm is executed by decreasing $i$ until
either an empty interval is found (i.e. $sp_i> ep_i$), or
$A[sp_1,ep_1]$ contains all pattern occurrences. In the former
case no pattern occurrences are found; in the latter case the
algorithm has found $occ = ep_1 - sp_1 +1$ pattern occurrences.

\subsection{Rank Query}
\label{sec:rankselwt}

Given a string $S[1,n]$, function $rank_x(S,i)$ returns the number
of times symbol $x$ appears in the prefix $S[1,i]$. Rank queries
are central to compressed indexing, so it is important to
understand how they are implemented and how much space/time they
need. We have two cases depending on the alphabet of $S$.

\smallskip \noindent {\bf Rank over Binary Sequences.} In this case
there exist simple and practical constant-time solutions using
$o(n)$ bits of space in addition to $S$ \cite{Mun96_tables}. We
cover only $rank_1$ as $rank_0(S,i) = i - rank_1(S,i)$. One of the
most efficient solutions in practice \cite{GGMNwea05} consists of
partitioning $S$ into blocks of size $s$, and storing explicit
answers for rank-queries done at block beginnings. Answering
$rank_1(S,i)$ then consists of summing two quantities: (1) the
pre-computed answer for the prefix of $S$ which ends at the
beginning of the block enclosing $S[i]$, plus (2) the {\em
relative} rank of $S[i]$ within its block. The latter is computed
via a byte-wise scanning of the block, using small precomputed
tables. This solution involves a space/time tradeoff related to
$s$, but nonetheless its query-time performance is rather
satisfactory already with 5\% space overhead on top of $S$.

\smallskip \noindent {\bf Rank over General Sequences.} Given a sequence
$S[1,n]$ over an alphabet of size $\sigma$, the {\em wavelet tree}
\cite{GGV03_high_order,GGV06_indexing_equals} is a perfect binary
tree of height $\Theta(\log \sigma)$, built on the alphabet
symbols, such that the root represents the whole alphabet and each
leaf represents a distinct alphabet symbol. If a node $v$
represents alphabet symbols in the range $\Sigma^v = [i,j]$, then
its left child $v_l$ represents $\Sigma^{v_l} = [i,\frac{i+j}{2}]$
and its right child $v_r$ represents $\Sigma^{v_r} =
[\frac{i+j}{2}+1,j]$. We associate to each node $v$ the
subsequence $S^v$ of $S$ formed by the symbols in $\Sigma^v$.
Sequence $S^v$ is not really stored at the node, but it is
replaced by a bit sequence $B^v$ such that $B^v[i]=0$ iff $S^v[i]$
is a symbol whose leaf resides in the left subtree of $v$.
Otherwise, $B^v[i]$ is set to 1.

The power of the wavelet tree is to reduce rank operations over
general alphabets to rank operations over a binary alphabet, so
that the rank-machinery above can be used in each wavelet-tree
node. Precisely, let us answer the query $rank_c(S,i)$.
We start from the root $v$ of the wavelet tree (with associated
vector $B^v$), and check which subtree encloses the queried
symbol $c$. If $c$ descends into the right subtree, we set $ i
\leftarrow rank_1(B^v,i)$ and move to the right child of $v$.
Similarly, if $c$ belongs to the left subtree, we set $i
\leftarrow rank_0(B^v,i)$ and go to the left child of $v$. We
repeat this until we reach the leaf that represents $c$, where the
current $i$ value is the answer to $rank_c(S,i)$. Since any
binary-rank takes $O(1)$ time, the overall rank operation takes
$O(\log \sigma)$ time.

We note that the wavelet tree can replace $S$ as well: to obtain
$S[i]$, we start from the root $v$ of the wavelet tree. If
$B^v[i]=0$, then we set $i \leftarrow rank_0(B^v, i)$ and go to
the left child. Similarly, if $B^v[i]=1$, then we set $ i
\leftarrow rank_1(B^v,i)$ and go to the right child. We repeat
this until we reach a leaf, where the symbol associated to the
leaf is the answer. Again, this takes $O(\log \sigma)$ time.

The wavelet tree requires comparable space to the original
sequence, as it requires $n \log\sigma~(1+o(1))$ bits of space. A
practical way to reduce the space occupancy to the zero-order
entropy of $S$ is to replace the balanced tree structure by the
Huffman tree of $S$. Now we have to follow the binary Huffman code
of a symbol to find its place in the tree. It is not hard to see
that the total number of bits required by such a tree is at most
$n (H_0(S)+1) + o(n\log\sigma)$ and the average time taken by rank
and access operations is $O(H_0(S))$, where $H_0$ is the zero-th
order empirical entropy of $S$ (see next section). This structure
is the key tool in our implementation of \ssa\ or \afindex\ 
(Section~\ref{sec:exper}).

\subsection{The $k$-th Order Empirical Entropy}
\label{sec:empirical}

The {\em empirical entropy} resembles the entropy defined in the
probabilistic setting (for example, when the input comes from a
Markov source), but now it is defined for any finite individual
string and can be used to measure the performance of compression
algorithms without any assumption on the input distribution
\cite{Man01_An_analysis}.

The empirical zero-order entropy of a text $T$ is defined
as
\begin{equation}
H_{0}(T)~=~ \sum_{c \in \Sigma}\frac{n_{c}}{n}\log
\frac{n}{n_{c}} ~, \label{eq1}
\end{equation}
where $n_{c}$ is the number of occurrences of symbol $c$ in $T$.
This definition extends to $k>0$ as follows. Let $\Sigma^k$ be the
set of all sequences of length $k$ over $\Sigma$. For any string
$w \in \Sigma^k$, called a {\em context} of size $k$, let $w_{T}$
be the string consisting of the concatenation of individual
symbols following $w$ in $T$. Then, the $k$-th order empirical
entropy of $T$ is defined as
\begin{equation}
H_{k}(T)~=~\frac{1}{n}\sum\limits_{w\in A^{k}}|w_{T}| \:
H_{0}\left(w_{T}\right). \label{eq2}
\end{equation}

The $k$-th order empirical entropy captures the dependence of
symbols upon their $k$-long context. For $k\geq 0$, $nH_{k}(T)$
provides a lower bound to the number of bits output by any
compressor that considers a context of size $k$ to encode each
symbol of $T$ (e.g. PPM-like compressors). Note that $0 \le H_k(T)
\le H_{k-1}(T) \le \ldots \le H_1(T) \le H_0(T) \le \log\sigma$.
Several compressed indexes achieve $O(nH_k(T^r))$ bits of space,
instead of $O(nH_k(T))$, as they work on the contexts {\em
following} (rather than preceding) the symbol to be encoded.
Nonetheless, we will not point out such a difference because one
can always work on the reversed text (and patterns) if necessary,
and also because both $k$-th order entropies differ by lower order
terms \cite{FM05}.

\subsection{The Burrows-Wheeler Transform}
\label{sec:bwt}

The  {\em Burrows-Wheeler Transform (BWT)} \cite{BW94_transform}
is a key tool in designing compressed full-text indexes. It is a
reversible permutation of $T$, which has the nice property of
putting together symbols followed by the same context. This
ensures that the permuted $T$ offers better compression
opportunities: a locally adaptive zero-order compressor is
able to achieve on this string the $k$-th order entropy of $T$
(recall Eq.~(\ref{eq2})). The BW-transform works as follows:

\begin{enumerate}
\item Create a conceptual matrix $M$, whose rows are cyclic shifts of $T$.

\item Sort the matrix rows lexicographically.

\item Define the last column of $M$ as the BWT of $T$, and call it $T^{bwt}$.
\end{enumerate}

There is a close relationship between matrix $M$ and the suffix
array $A$ of text $T$, because when we lexicographically sort the
rows, we are essentially sorting the suffixes of $T$ (recall
indeed that $t_n=\$$ is smaller than any other alphabet symbol).
Specifically, $A[i]$ points to the suffix of $T$ which prefixes
the $i$-th row of $M$. Hence, another way to describe $T^{bwt}$ is
to concatenate the symbols that precede each suffix of $T$ in the
order listed by $A$, that is, $T^{bwt} ~=~
t_{A[1]-1}~t_{A[2]-1}\ldots t_{A[n]-1}$, where we assume that $t_0
= t_n$.

Given the way matrix $M$ has been built, all columns of $M$ are
permutations of $T$. So the first and last column of $M$ are
indeed one a permutation of the other. The question is how to
map symbols in the last column $T^{bwt}$ to symbols in the first
column. It is easy to see \cite{BW94_transform} that
occurrences of equal symbols preserve their relative order in the
last and the first columns of $M$. Thus the $j$-th occurrence of a
symbol $c$ within $T^{bwt}$ corresponds to the $j$-th occurrence
of $c$ in the first column. If $c=T^{bwt}[i]$, then we have that
$j = rank_c(T^{bwt},i)$ in the last column; whereas in the first
column, where the symbols are sorted alphabetically, the $j$-th
occurrence of $c$ is at position $C[c]+j$, where $C[c]$ counts the
number of occurrences in $T$ of symbols smaller than $c$. By
plugging one formula in the other we derive the so called {\em
Last-to-First column mapping} (or, LF-mapping):
$LF(i)=C[c]+rank_{c}(T^{bwt},i)$. We talk about LF-mapping because
the symbol $c= T^{bwt}[i]$ is located in the first column of $M$
at position $LF(i)$.

The LF-mapping allows one to navigate $T$ backwards: if
$t_k=T^{bwt}[i]$, then $t_{k-1}=T^{bwt}[LF(i)]$ because row
$LF(i)$ of $M$ starts with $t_k$ and thus ends with $t_{k-1}$. As
a result we can reconstruct $T$ backwards by starting at the first
row, equal to $\$T$, and repeatedly applying $LF$ for $n$ steps.

\section{Compressed Indexes}
\label{sec:indexes}

As explained in the Introduction, compressed indexes provide a
viable alternative to classical indexes that are parsimonious
in space and efficient in query time. They have undergone
significant development in the last years, so that we count now in
the literature many solutions that offer a plethora of space-time
tradeoffs \cite{NMacmcs06}. In theoretical terms, the most succinct
indexes achieve $nH_k(T)+o(n\log\sigma)$ bits of space, and for a
fixed $\epsilon>0$, require $O(m\log\sigma)$ counting time,
$O(\log^{1+\epsilon} n)$ time per located occurrence,  and
$O(\ell\log\sigma+\log^{1+\epsilon} n)$ time to extract a
substring of $T$ of length $\ell$.\footnote{These locating and
extracting complexities are better than those reported in
\cite{FMMN-TALG}, and can be obtained by setting the sampling step
to $\frac{\log^{1+\epsilon}n}{\log \sigma}$.} This is a surprising
result because it shows that whenever $T[1,n]$ is compressible it
can be indexed into smaller space than its plain
form and still offer search capabilities in efficient time.

In the following we review the most competitive compressed indexes
for which there is an implementation we are aware of. We will
review the FM-index family, which builds on the BWT and backward
searching; Sadakane's Compressed Suffix Array (CSA), which is
based on compressing the suffix array via a so-called $\Psi$
function that captures text regularities; and the LZ-index, which
is based on Lempel-Ziv compression. All of them are self-indexes
in that they include the indexed text, which therefore may be
discarded.

\subsection{The FM-index Family}\label{sec:FM-index}

The FM-index is composed of a compressed representation of
$T^{bwt}$ plus auxiliary structures for efficiently computing
generalized rank queries on it. The main idea \cite{FM05} is to
obtain a text index from the BWT and then use backward searching
for identifying the pattern occurrences
(Sections~\ref{sec:bsearch} and \ref{sec:bwt}). Several variants
of this algorithmic scheme do exist
\cite{FM01,FM05,MNnjc05,FMMN-TALG} which induce several time/space
tradeoffs for the counting, locating, and extracting operations.

\medskip \noindent {\bf Counting.} The counting procedure takes
a pattern $P$ and obtains the interval $A[sp,ep]$ of text suffixes
prefixed by it (or, which is equivalent, the interval of rows of
the matrix $M$ prefixed by $P$, see Section~\ref{sec:bwt}). Fig.
\ref{fig:getrow} gives the pseudocode to compute $sp$ and $ep$.

\begin{figure}[tbh]
\hrule \smallskip
\textbf{Algorithm} FM-count($P_{1,m}$)
\par
$i\leftarrow m,~sp\leftarrow 1,~ep\leftarrow n;$%
\par
\textbf{while} $((sp\leq ep)\ \textbf{and}\ (i\geq 1))$\ \textbf{do}
\par
\qquad $c\leftarrow p_i$;
\par
\qquad $sp\leftarrow C[c]+rank_c(T^{bwt},sp-1)+1;$%
\par
\qquad $ep\leftarrow C[c]+rank_c(T^{bwt},ep);$%
\par
\qquad $i\leftarrow i-1;$%
\par
\textbf{if} $(sp > ep)$\ \textbf{then return} ``no occurrences of $P$''
    \textbf{else return} $\langle sp,ep \rangle$;
\smallskip \hrule
\caption{Algorithm to get the interval $A[sp,ep]$ of text suffixes
prefixed by $P$, using an FM-index.} 
\label{fig:getrow}
\end{figure}

The algorithm is correct: Let $[sp_{i+1},ep_{i+1}]$ be the range
of rows in $M$ that start with $P_{i+1,m}$, and we wish to know which
of those rows are preceded by $p_i$. These correspond precisely to
the occurrences of $p_i$ in $T^{bwt}[sp_{i+1},ep_{i+1}]$. Those
occurrences, mapped to the first column of $M$, form a
(contiguous) range that is computed with a rationale similar to
that for $LF(\cdot)$ in Section~\ref{sec:bwt}, and thus via a just
two rank operations.

\medskip \noindent {\bf Locating.} Algorithm in Fig. \ref{fig:locate} obtains the
position of the suffix that prefixes the $i$-th row of $M$. The
basic idea is to logically mark a suitable set of rows of $M$, and
keep for each of them their position in $T$ (that is, we store the
corresponding $A$ values). Then, FM-locate($i$) scans the text $T$ backwards 
using the LF-mapping until a marked row $i^{\prime }$
is found, and then it reports $A[i^{\prime }]+t,$ where $t$ is the
number of backward steps used to find such $i^{\prime }$. To
compute the position of all occurrences of a pattern $P$, it is
thus enough to call FM-locate($i$) for every $sp \le i \le ep$.

\begin{figure}[tbh]
\hrule \smallskip
\textbf{Algorithm} FM-locate($i$)
\par
$i^{\prime }\leftarrow i,t\leftarrow 0$;
\par
\textbf{while} $A[i^{\prime }]$ is not explicitly stored \textbf{do}
\par
\qquad $i^{\prime }\leftarrow LF(i^{\prime })$;
\par
\qquad $t\leftarrow t+1$;
\par
\textbf{return} $A[i^{\prime }]\ +t$;
\smallskip \hrule
\caption{Algorithm to obtain $A[i]$ using an FM-index.}
\label{fig:locate}
\end{figure}

The sampling rate of $M$'s rows, hereafter denoted by $s_A$, is a
crucial parameter that trades space for query time. Most FM-index
implementations mark all the $A[i]$ that are a multiple of $s_A$,
via a bitmap $B[1,n]$. All the marked $A[i]$s are stored
contiguously in suffix array order, so that if $B[i]=1$ then one
finds the corresponding $A[i]$ at position $rank_1(B,i)$ in that
contiguous storage. This guarantees that at most $s_A$ LF-steps
are necessary for locating the text position of any occurrence.
The extra space is $\frac{n\log n}{s_A} + n + o(n)$ bits.

A way to avoid the need of bitmap $B$ is to choose a symbol $c$
having some suitable frequency in $T$, and then store $A[i]$ if
$T^{bwt}[i]=c$ \cite{FM01}. Then the position of $A[i]$ in the
contiguous storage is $rank_c(T^{bwt},i)$, so no extra space is
needed other than $T^{bwt}$. In exchange, there is no guarantee of
finding a marked cell after a given number of steps.

\medskip \noindent {\bf Extracting.} The same text sampling mechanism
used for locating permits extracting text substrings. Given $s_A$,
we store the positions $i$ such that $A[i]$ is a multiple of $s_A$
now in the text order (previously we followed the $A$-driven
order). To extract $T_{l,r}$, we start from the first sample that
follows the area of interest, that is, sample number $d = \lceil
(r+1)/s_A \rceil$. From it we obtain the desired text
backwards with the same mechanism for inverting the BWT (see 
Section~\ref{sec:bwt}), here starting with the value $i$ stored
for the $d$-th sample. We need at most $s_A + r - l+1$
applications of the LF-step.

\subsection{Implementing the FM-index}

All the query complexities are governed by the
time required to obtain $C[c]$, $T^{bwt}[i]$, and
$rank_c(T^{bwt},i)$ (all of them implicit in $LF$ as well). While
$C$ is a small table of $\sigma\log n$ bits, the other two are
problematic. Counting requires up to $2m$ calls to $rank_c$,
locating requires $s_A$ calls to $rank_c$ and $T^{bwt}$, and
extracting $\ell$ symbols requires $s_A + \ell$ calls to $rank_c$
and $T^{bwt}$. In what follows we briefly comment on the solutions
adopted to implement those basic operations.

The original FM-index implementation ({\em FM-index} \cite{FM01})
compressed $T^{bwt}$ by splitting it into blocks and using
independent zero-order compression on each block. Values of
$rank_c$ are precomputed for all block beginnings, and the rest of
the occurrences of $c$ from the beginning of the block to any
position $i$ are obtained by sequentially decompressing the block.
The same traversal finds $T^{bwt}[i]$. This is very
space-effective: It approaches in practice the $k$-th order
entropy because the partition into blocks takes advantage of the
local compressibility of $T^{bwt}$. On the other hand, the time to
decompress the block makes computation of $rank_c$ relatively
expensive. For locating, this implementation marks the BWT
positions where some chosen symbol $c$ occurs, as explained above.

A very simple and effective alternative to represent $T^{bwt}$ has
been proposed with the {\it Succinct Suffix Array} ({\it SSA})
\cite{FMMN-TALG,MNnjc05}. It uses a Huffman-shaped wavelet tree,
plus the marking of one out-of $s_A$ text positions for locating
and extracting. The space is $n(H_0(T)+1)+o(n\log\sigma)$ bits,
and the average time to determine $rank_c(T^{bwt},i)$ and
$T^{bwt}[i]$ is $O(H_0(T)+1)$. The space bound is not
appealing because of the zero-order compression, but the relative
simplicity of this index makes it rather fast in practice. In
particular, it is an excellent option for DNA text, where the
$k$-th order compression is not much better than the zero-th order
one, and the small alphabet makes $H_0(T) \leq \log\sigma$ small
too.

The {\em Run-Length FM-index} ({\em RLFM}) \cite{MNnjc05} has been
introduced to achieve $k$-th order compression by applying {\em
run-length compression} to $T^{bwt}$ prior to building a wavelet
tree on it. The BWT generates long runs of identical symbols on
compressible texts, which makes the RLFM an interesting
alternative in practice. The price is that the mappings from the
original to the run-length compressed positions slow down the
query operations a bit, in comparison to the SSA.

\subsection{The Compressed Suffix Array (CSA)}
\label{sec:csa}

The compressed suffix array ({\em CSA}) was not originally a
self-index, and required $O(n\log\sigma)$ bits of space
\cite{GV05}. Sadakane \cite{Sad03,Sad02_succinct} then proposed a
variant which is a self-index and achieves high-order compression.

The CSA represents the suffix array $A[1,n]$ by a sequence of
numbers $\psi(i)$, such that $A[\psi(i)]=A[i]+1$. It is not hard to see
\cite{Sad03} that  $\psi$ is piecewise monotone increasing
in the areas of $A$ where the suffixes start with the same symbol.
In addition, there are long runs where $\psi(i+1) = \psi(i)+1$,
and these runs can be mapped one-to-one to the runs in $T^{bwt}$
\cite{NMacmcs06}. These properties permit a compact representation
of $\psi$ and its fast access. Essentially, we differentially
encode $\psi(i) -\psi(i-1)$, run-length encode the long runs
of 1's occurring over those differences, and for the rest use 
an encoding favoring small numbers. Absolute samples are
stored at regular intervals to permit the efficient decoding of
any $\psi(i)$. The sampling rate (hereafter denoted by $s_\psi$)
gives a space/time tradeoff for accessing and storing $\psi$. In
\cite{Sad03} it is shown that the index requires
$O(nH_0(T)+n\log\log\sigma)$ bits of space. The analysis has been
then improved in \cite{NMacmcs06} to $nH_k(T)+O(n\log\log\sigma)$
for any $k \le \alpha \log_\sigma n$ and constant $0<\alpha<1$. 

\medskip \noindent {\bf Counting.} The original CSA \cite{Sad03} used the
classical binary searching to count the number of pattern
occurrences in $T$. The actual implementation, proposed in
\cite{Sad02_succinct}, uses backward searching
(Section~\ref{sec:bsearch}): $\psi$ is used to obtain
$\langle sp_i,ep_i\rangle$ from $\langle sp_{i+1},ep_{i+1}\rangle$ in 
$O(\log n)$ time, for
a total of $O(m\log n)$ counting time. Precisely, let
$A[sp_i,ep_i]$ be the range of suffixes $A[j]$ that start with
$p_i$ and such that $A[j]+1$ ($= A[\psi(j)]$) starts with
$P_{i+1,m}$. The former is equivalent to the condition
$[sp_i,ep_i] \subseteq [C[p_i]+1,C[p_i+1]]$. The latter is
equivalent to saying that $sp_{i+1} \le \psi(j) \le ep_{i+1}$. Since
$\psi(i)$ is monotonically increasing in the range $C[p_i] < j \le
C[p_i+1]$ (since the first characters of suffixes in
$A[sp_i,ep_i]$ are the same), we can binary search this interval to
find the range $[sp_i,ep_i]$. Fig. \ref{fig:csa_alg} shows the
pseudocode for counting using the CSA.

\begin{figure}[tbh]
\hrule \smallskip
\textbf{Algorithm} CSA-count($P_{1,m}$)
\par
$i\leftarrow m,~sp\leftarrow 1,~ep\leftarrow n;$%
\par
\textbf{while} $((sp\leq ep)\ \textbf{and} (i\geq 1))$\ \textbf{do}
\par
\qquad $c\leftarrow p_i$;
\par
\qquad $\langle sp,ep\rangle \leftarrow \langle \min,\max\rangle ~\{j \in [C[c]+1,C[c+1]], \psi(j) \in [sp,ep]\}$;
\par
\qquad $i\leftarrow i-1;$
\par
\textbf{if} $(ep<sp)$\ \textbf{then return} ``no occurrences of
$P$'' \textbf{else return} $\langle sp,ep\rangle$;
\smallskip \hrule
\caption{Algorithm to get the interval $A[sp,ep]$ prefixed by $P$,
using the CSA. The $\langle \min,\max\rangle$ interval is obtained via binary
search.} 
\label{fig:csa_alg}
\end{figure}

\medskip \noindent {\bf Locating.} Locating is similar to the
FM-index, in that the suffix array is sampled at regular intervals
of size $s_A$. However, instead of using the LF-mapping to
traverse the text backwards, this time we use $\psi$ to traverse
the text forward, given that $A[\psi(i)] = A[i]+1$. This points
out an interesting duality between the FM-index and the CSA. Yet,
there is a fundamental difference: function $LF(\cdot)$ is
implicitly stored and calculated on the fly over $T^{bwt}$, while
function $\psi(\cdot)$ is explicitly stored. The way these
functions are calculated/stored makes the CSA a better alternative
for large alphabets.

\medskip\noindent {\bf Extracting.} Given $C$ and $\psi$, we can
obtain $T_{A[i],n}$ symbolwise from $i$, as follows. The first
symbol of the suffix pointed to by $A[i]$, namely $t_{A[i]}$, is
the character $c$ such that $C[c] < i \le C[c+1]$, because all the
suffixes $A[C[c]+1],\ldots,A[C[c+1]]$ start with symbol $c$. Now,
to obtain the next symbol, $t_{A[i]+1}$, we compute $i'=\psi(i)$
and use the same procedure above to obtain $t_{A[i']}=t_{A[i]+1}$,
and so on. The binary search in $C$ can be avoided by representing
it as a bit vector $D[1,n]$ such that $D[C[c]]=1$, thus $c =
rank_1(D,i)$.

Now, given a text substring $T_{l,r}$ to extract, we must first
find the $i$ such that $l = A[i]$ and then we can apply the
procedure above. Again, we sample the
text at regular intervals by storing the $i$ values such that
$A[i]$ is a multiple of  $s_A$. To extract $T_{l,r}$ we actually
extract $T_{\lfloor l/s_A \rfloor \cdot s_A,r}$, so as to start
from the preceding sampled position. This takes $s_A + r - l+1$
applications of $\psi$.

\subsection{The Lempel-Ziv Index}
\label{sec:lzi}

The {\em Lempel-Ziv index (LZ-index)} is a compressed self-index
based on a Lempel-Ziv partitioning of the text. There are several
members of this family \cite{Navjda03,ANS06,FM05}, we focus on
the version described in \cite{Navjda03,ANS06} and available in
the Pizza\&Chili site. This index uses LZ78 parsing
\cite{ziv78compression} to generate a partitioning of $T_{1,n}$
into $n'$ {\em phrases}, $T = Z_1,\ldots,Z_{n'}$. These phrases
are all different, and each phrase $Z_i$ is formed by appending a
single symbol to a previous phrase $Z_j$, $j<i$ (except for a
virtual empty phrase $Z_0$). Since it holds $Z_i = Z_j\cdot c$,
for some $j<i$ and $c\in\Sigma$, the set is prefix-closed. We can
then build a trie on these phrases, called LZ78-trie, which
consists of $n'$ nodes, one per phrase.

The original LZ-index \cite{Navjda03} is formed by (1) the LZ78
trie; (2) a trie formed with the reverse phrases $Z_i^r$, called
the reverse trie; (3) a mapping from phrase identifiers $i$ to the
LZ78 trie node that represents $Z_i$; and (4) a similar mapping to
$Z_i^r$ in the reverse phrases. The tree shapes in (1) and (2) are
represented using parentheses and the encoding proposed in
\cite{MR97_parentheses} so that they take $O(n')$ bits and
constant time to support various tree navigation operations. Yet,
we must also store the phrase identifier in each trie node, which
accounts for the bulk of the space for the tries. Overall, we have
$4n'\log n'$ bits of space, which can be bounded by $4n H_k(T) +
o(n\log\sigma)$ for $k=o(\log_\sigma n)$ \cite{NMacmcs06}. This
can be reduced to $(2+\epsilon)n H_k(T) + o(n\log\sigma)$ by
noticing\footnote{This kind of space reductions is explored in
\cite{ANS06}. The one we describe here is not reported there, but
has been developed by those authors for the {\em Pizza\&Chili}
site and will be included in the journal version of \cite{ANS06}.}
that the mapping (3) is essentially the inverse permutation of the
sequence of phrase identifiers in (1), and similarly (4) with (2).
It is possible to represent a permutation and its inverse using
$(1+\epsilon)n'\log n'$ bits of space and access the inverse
permutation in $O(1/\epsilon)$ time
\cite{MRRR03_succinct_permutations}.

An occurrence of $P$ in $T$ can be found according to one of the
following situations:
\begin{enumerate}

\item $P$ lies within a phrase $Z_i$. Unless the occurrence is a suffix of
$Z_i$, since $Z_i = Z_j\cdot c$, $P$ also appears within $Z_j$,
which is the parent of $Z_i$ in the LZ78 trie. A search for $P^r$
in the reverse trie finds all the phrases that have $P$ as a
suffix. Then the node mapping permits, from the phrase identifiers
stored in the reverse trie, to reach their corresponding LZ78
nodes. All the subtrees of those nodes are occurrences.

\item $P$ spans two consecutive phrases. This means that, for some $j$,
$P_{1,j}$ is a suffix of some $Z_i$ and $P_{j+1,m}$ is a prefix of
$Z_{i+1}$. For each $j$, we search for $P_{1,j}^r$ in the reverse
trie and $P_{j+1,m}$ in the LZ78 trie, choosing the smaller
subtree of the two nodes we arrived at. If we choose the
descendants of the reverse trie node for $P_{1,j}^r$, then for
each phrase identifier $i$ that descends from the node, we check
whether $i+1$ descends from the node that corresponds to
$P_{j+1,m}$ in the LZ78 trie. This can be done in constant time by
comparing preorder numbers.

\item $P$ spans three or more nodes. This implies that some phrase is
completely contained in $P$, and since all phrases are different, there are
only $O(m^2)$ different phrases to check, one per substring of $P$. Those are
essentially verified one by one.
\end{enumerate}

Notice that the LZ-index carries out counting and locating
simultaneously, which renders the LZ-index not competitive for
counting alone. Extracting text is done by traversing the LZ78
paths upwards from the desired phrases, and then using mapping (3)
to continue with the previous or next phrases. The LZ-index is
very competitive for locating and extracting.

\subsection{Novel Implementations}

We introduce two novel compressed index implementations in this paper. Both
are variants of the FM-index family. The first one is interesting because it
is a re-engineering of the first reported implementation of a self-index
\cite{FM01}. The second is relevant because it implements the self-index
offering the best current theoretical space/time guarantees. It is fortunate,
as it does not always happen, that theory and practice marry well and this
second index is also relevant in the {\em practical} space/time tradeoff map.

\subsubsection{The FMI-2}
\label{sec:fmin}

As the original FM-index \cite{FM01}, the {\em FMI-2} adopts a 
two-level bucketing scheme for implementing efficient {\em rank}
and {\em access} operations onto $T^{bwt}$. In detail, string
$T^{bwt}$ is partitioned into {\em buckets} and {\em superbuckets}:
a bucket consists of $lb$ symbols, a superbucket consists of $lsb$
buckets. Additionally, the FMI-2 maintains two tables: Table $T_{sb}$
stores, for each superbucket and for each symbol $c$, the number of
occurrences of $c$ before that superbucket in $T^{bwt}$; table $T_b$
stores, for each bucket and for each symbol $c$, the number of
occurrences of $c$ before that bucket and up to the beginning of its
superbucket. In other words, $T_{sb}$ stores the value of the
ranking function up to the beginning of superbuckets; whereas $T_b$
stores the ranking function up to the beginning of buckets and {\em
relative} to their enclosing superbuckets. Finally, every bucket is
individually compressed using the sequence of zero-order
compressors: MTF, RLE, Huffman (as in \bzip). This compression
strategy does not guarantee that the space of FMI-2 is bounded by
the $k$th order entropy of $T$. Nevertheless, the practical
performance is close to the one achievable by the best known
compressors, and can be traded by tuning parameters $lb$ and $lsb$.

The main difference between the original FM-index and the novel FMI-2 relies in
the strategy adopted to select the rows/positions of $T$ which are
explicitly stored. The FMI-2 marks logically and uniformly the text $T$
by adding a special symbol every $s_A$ symbols of the original text.
This way, all of the $M$'s rows that start with that special symbol
are contiguous, and thus their positions can be stored and accessed
easily.

\no{}{
Given these auxiliary data structures, we detail now how the rank
operation is implemented, and then sketch counting and locating
operations. The extraction of text substrings is implemented mainly
as detailed in Section~\ref{sec:FM-index}, with attention to deal
with the presence of the special symbol.

$Rank_c(T^{bwt},i)$ is implemented by summing up three quantities
which totally account for the number of occurrences of $c$ in the
prefix $T^{bwt}[1,i]$. Precisely, we decompress and scan the bucket
that contains $T^{bwt}[i]$ in order to count the number of
occurrences of $c$ from the beginning of the bucket and up to
$T^{bwt}[i]$. The number of occurrences from the beginning of
$T^{bwt}$ is eventually obtained by accessing tables $T_{sb}$ and
$T_{b}$. It goes without saying that the values of $lb$ and $lsb$
provide a tradeoff between compression ratio and efficiency of rank,
and thus impact onto the performance of all operations executed on
the FMI-2.
}

The count algorithm is essentially a backward search (Algorithm
\ref{fig:getrow}), modified to take into account the presence of
special symbols added to the indexed text. To search for a pattern
$P_{1,p}$, the FMI-2 actually searches for $min\{p-1,s_A\}$ patterns
obtained by inserting the special symbols in $P$ at each $s_A$-th position, 
and searches for the pattern $P$ itself. This
search is implemented {\em in parallel} over all patterns above by
exploiting the fact that, at any step $i$, we have to search either
for $P_{p-i}$ or for the special symbol. As a result, the overall
search cost is quadratic in the pattern length, and the output is
now a set of at most $p$ ranges of rows.

Therefore, the FMI-2 is slower in counting than the original FM-index, but 
locating is faster, and this is crucial because this latter operation is 
usually the bottleneck of compressed indexes. Indeed the locate algorithm
proceeds for at most $s_A$ phases. Let $S_0$ be the range of rows to
be located, eventually identified via a count operation. At a
generic phase $k$, $S_k$ contains the rows that may be reached in
$k$ backward steps from the rows in $S_0$. $S_k$ consists of a {\em
set} of ranges of rows, rather than a single range. To maintain the
invariant, the algorithm picks up a range of $S_k$, say $[a,b]$, and
determines the $z\leq|\Sigma|$ distinct symbols that occur in the
substring $T^{bwt}_{a,b}$ via two bucket scans and some accesses to
tables $T_{sb}$ and $T_{b}$. Then it executes $z$ backward steps,
one per such symbols, thus determining $z$ new ranges of rows (to be
inserted in $S_{k+1}$) which are at distance $k+1$ from the rows in
$S_0$. The algorithm cycles over all ranges of $S_k$ to form the new
set $S_{k+1}$. Notice that if the rows of a range start with the
special symbol, their positions in the indexed text are explicitly
stored, and can be accessed in constant time. Then, the position of
the corresponding rows in $S_0$ can be inferred by summing $k$ to
those values. Notice that this range can be dropped from $S_k$.
After no more than $s_A$ phases the set $S_k$ will be empty.

\subsubsection{The Alphabet-Friendly FM-index}
\label{sec:afmi}

The {\it Alphabet-Friendly FM-index} ({\it AF-index})
\cite{FMMN-TALG} resorts to the definition of $k$-th order entropy in
Eq.~(\ref{eq2}), by encoding each substring $w_T$ up to its
zero-order entropy. Since all the $w_T$ are contiguous in
$T^{bwt}$ (regardless of which $k$ value we are considering), it
suffices to split $T^{bwt}$ into blocks given by the $k$-th order
contexts, for any desired $k$, and to use a Huffman-shaped wavelet
tree (see Section~\ref{sec:rankselwt}) to represent each such block. 
In addition, we need all
$rank_c$ values precomputed for every block beginning, as the
local wavelet trees can only answer $rank_c$ within their blocks.
In total, this achieves $nH_k(T)+o(n\log\sigma)$ bits, for moderate
and fixed $k \le \alpha \log_\sigma n$ and $0<\alpha<1$. 
Actually the AF-index does better, by splitting $T^{bwt}$
in an {\em optimal way}, thus guaranteeing that the space bound above holds 
simultaneously for every $k$. This is done by resorting to the idea of 
{\it compression boosting} \cite{FM04_compression_boosting,GS03_Optimal}.

The compression booster finds the optimal partitioning of $T^{bwt}$ into
$t$ nonempty blocks, $s_1,\ldots,s_t$, assuming that each block $s_j$ 
will be represented using $|s_j| H_0(s_j)+f(|s_j|)$ bits of space, where
$f(\cdotp)$ is a nondecreasing concave function supplied as a parameter. 
Given that the partition is optimal, it can be shown that the resulting
space is upper bounded by $nH_k + \sigma^k f(n/\sigma^k)$ bits {\em
simultaneously for every $k$}. That is, the index is not built for any
specific $k$.

As explained, the AF-index represents each block $s_j$ by means of a 
Huffman-shaped wavelet tree $wt_j$, which will take at most 
$|s_j|(H_0(s_j)+ 1)+\sigma\log n$ bits. The last
term accounts for the storage of the Huffman code. In addition, for each block 
$j$ we store an array $C_j[c]$, which tells the $rank_c$ values up to block
$j$. This accounts for other $\sigma\log n$ bits per block. Finally, we need
a bitmap $R[1,n]$ indicating the starting positions of the $t$ blocks in
$T^{bwt}$. Overall, the formula giving the excess of storage over the entropy 
for block $j$ is $f(|s_j|) = 2|s_j|+2\sigma\log n$.

%
%
%

To carry out any operation at position $i$, we start by computing the
block where position $i$ lies, $j = rank_1(R,i)$, and the starting position
of that block, $i' = select_1(R,j)$. (This tells the position of the $j$-th 1
in $R$. As it is a sort of inverse of $rank$, it is computed by binary search
over $rank$ values.) Hence $T^{bwt}[i] = s_j[i'']$, where
$i''=i-i'+1$ is the offset of $i$ within block $j$.
Then, the different operations are carried out as follows.

\begin{itemize}
\item For counting, we use the algorithm of Fig. \ref{fig:getrow}. In this 
case, we have $rank_c(T^{bwt},i)\hspace{-3.42pt}=C_j[c]+rank_c(s_j,i'')$, where the latter is
computed using the wavelet tree $wt_j$ of $s_j$.
\item For locating, we use the algorithm of Fig. \ref{fig:locate}. In this 
case, we have $c=T^{bwt}[i]=s_j[i'']$. To compute $s_j[i'']$, we also use the 
wavelet tree $wt_j$ of $s_j$.
\item For extracting, we proceed similarly as for locating, as explained in
Section~\ref{sec:FM-index}.
\end{itemize}

As a final twist, $R$ is actually stored using $2\sqrt{nt}$ rather than $n$
bits. We cut $R$ into $\sqrt{nt}$ chunks of length $\sqrt{n/t}$. There are
at most $t$ chunks which are not all zeros. Concatenating them all requires
only $\sqrt{nt}$ bits. A second bitmap of length $\sqrt{nt}$ indicates whether
each chunk is all-zero or not. It is easy to translate rank/select operations
into this representation.

\section{The Pizza\&Chili Site}
\label{sec:site}

The {\em Pizza\&Chili} site has two mirrors: one in Chile ({\footnotesize \tt
http://pizzachili.dcc.uchile.cl}) and one in Italy ({\footnotesize \tt
http://pizzachili.di.unipi.it}). 
Its ultimate goal is to push towards the technology transfer of this
fascinating algorithmic technology lying at the crossing point of
data compression and data structure design. In order to achieve
this goal, the {\em Pizza\&Chili} site offers publicly available
and highly tuned implementations of various compressed indexes.
The implementations follow a suitable C/C++ API of functions which
should, in our intention, allow any programmer to plug easily the
provided compressed indexes within his/her own software. The site
also offers a collection of texts for experimenting with and
validating the compressed indexes. In detail, it offers three
kinds of material:

\begin{itemize}
\item A set of compressed indexes which are able to support
the search functionalities of classical full-text indexes
(e.g., substring searches), but requiring succinct
space occupancy and offering, in addition, some text access
operations that make them useful within text retrieval and
data mining software systems.

\item A set of text collections of various types and sizes
useful to test experimentally the available (or new) compressed
indexes. The text collections have been selected to form a
representative sample of different applications where indexed text
searching might be useful. The size of these texts is large enough
to stress the impact of data compression over memory usage and CPU
performance. The goal of experimenting with this testbed is to
conclude whether, or not, compressed indexing is beneficial over
uncompressed indexing approaches, like suffix trees and suffix
arrays. And, in case it is beneficial, which compressed index is
preferable according to the various applicative scenarios
represented by the testbed.

\item Additional material useful to experiment with compressed
indexes, such as scripts for their automatic validation and
efficiency test over the available text collections.
\end{itemize}

The {\em Pizza\&Chili} site hopes to mimic the success and impact
of other initiatives, such as {\em data-compression.info} and the
{\em Calgary} and {\em Canterbury} corpora, just to cite a few.
Actually, the {\em Pizza\&Chili} site is a mix, as it offers both
software and testbeds. Several people have already contributed to
make this site work and, hopefully, many more will contribute to
turn it into a reference for all researchers and software
developers interested in experimenting and developing the
compressed-indexing technology. The API we propose is thus
intended to ease the deployment of this technology in real
software systems, and to provide a reference for any researcher
who wishes to contribute to the {\em Pizza\&Chili} repository with
his/her new compressed index.

\subsection{Indexes}
\label{sub:indexes}

The {\em Pizza\&Chili} site provides several index
implementations, all adhering to a common API. All indexes, except
CSA and LZ-index, are built through the deep-shallow algorithm of
Manzini and Ferragina \cite{FM02_Lightweight} which constructs the
Suffix Array data structure using little extra space and is fast 
in practice.

\begin{itemize}
\item The Suffix Array \cite{Manber_Myers93} is a plain
implementation of the classical index (see Section
\ref{sec:classic}), using either $n \log n$ bits of space or
simply $n$ computer integers, depending on the version. This was
implemented by Rodrigo Gonz\'alez.


\item The SSA \cite{FMMN-TALG,MNnjc05} uses a Huffman-based
wavelet tree over the string $T^{bwt}$ (Section
\ref{sec:FM-index}). It achieves zero-order entropy in space with
little extra overhead and striking simplicity. It was implemented
by Veli M\"akinen and Rodrigo Gonz\'alez.

\item The AF-index \cite{FMMN-TALG} combines compression
boosting \cite{FGMS-JACM} with the above wavelet tree data
structure (Section \ref{sec:afmi}). It achieves high-order
compression, at the cost of being more complex than SSA. It was
implemented by Rodrigo Gonz\'alez.

\item The RLFM \cite{MNnjc05} is an improvement over the SSA (Section \ref{sec:FM-index}),
which exploits the equal-letter runs of the BWT to achieve $k$-th
order compression, and in addition uses a Huffman-shaped wavelet tree. 
It is slightly larger than the AF-index. It was implemented by 
Veli M\"akinen and Rodrigo Gonz\'alez.

\item The FMI-2 (Section \ref{sec:fmin}) is
an engineered implementation of the original FM-index \cite{FM01},
where a different sampling strategy is designed in order to
improve the performance of the locating operation. It was
implemented by Paolo Ferragina and Rossano Venturini.

\item The CSA \cite{Sad03,Sad02_succinct} is the variant using 
backward search (Section \ref{sec:csa}). It achieves high-order
compression and is robust for large alphabets. It was implemented
by Kunihiko Sadakane and adapted by Rodrigo Gonz\'alez to adhere
the API of the {\em Pizza\&Chili} site. To construct the suffix
array, it uses the {\em qsufsort} by Jesper Larsson and Kunihiko
Sadakane \cite{LS99_qsuf}.

\item The LZ-index \cite{Navjda03,ANS06} is a compressed index
based on LZ78 compression (Section \ref{sec:lzi}), implemented by Diego
Arroyuelo and Gonzalo Navarro. It achieves high-order compression, yet with
relatively large constants. It is slow for counting but very competitive for
locating and extracting.
\end{itemize}

These implementations support any byte-based alphabet of size up
to 255 symbols: one symbol is automatically reserved by the
indexes as the terminator ``\$''.

In the following two sections we are going to explain the
implementation of \FMInew\ and \afindex.

\no{}{
\subsection{Programming Interface}
\label{exp:interface}

The {\em Pizza\&Chili} indexes are used through an API written in
C/C++ language. This API is sketched below. We use {\tt uchar} for
denoting unsigned char and {\tt ulong} for unsigned long. The
interface assumes that each text symbol is represented with one
byte. The integer {\tt e} returned by any procedure indicates an
error code, if it is different of zero. The error message can be
retrieved by calling the procedure {\tt char *error\_index(e)}.
Text and pattern indexes start at zero.

\begin{itemize}
\item {\bf Error management}
\begin{itemize}
\item {\tt char *error\_index (int e)} returns a string describing
          the error associated with $e$. The string must not be
          freed, and it will be overwritten with subsequent calls.
\end{itemize}
\item {\bf Building the index}
\begin{itemize}
\item {\tt int build\_index (uchar *text, ulong length, char *build\_options,
void} {\tt **index)} creates an index from {\tt
text[0..length-1]}. Note that the index is an opaque data type
(i.e. it uses {\tt void}-type). Build options must be passed in
string {\tt build\_options}, whose syntax depends on the index and
is described in its accompanying documentation. The index must
always work with some default parameters, if {\tt build\_options}
is {\tt NULL}. The returned index is ready to be queried.

\item {\tt int save\_index (void *index, char *filename)}
saves {\tt index} on disk by using single or multiple files, using
proper extensions.

\item {\tt int load\_index (char *filename, void **index)}
loads {\tt index} from one or more files named {\tt filename} with
proper extensions.

\item {\tt int free\_index (void *index)} frees the memory occupied by
{\tt index}.

\item {\tt int index\_size(void *index, ulong *size)} tells the memory
occupied by {\tt index} in bytes. This must be the internal memory
the index needs to operate.

\end{itemize}

\item {\bf Querying the index}
\begin{itemize}
\item {\tt int count (void *index, uchar *pattern, ulong length, ulong
*numocc)} writes in {\tt numocc} the number of occurrences of the
substring {\tt pattern[0..length-1]} found in the text indexed by
{\tt index}.

\item {\tt int locate (void *index, uchar *pattern, ulong length, ulong **occ, ulong *numocc)}
writes in {\tt numocc} the number of occurrences of {\tt
pattern[0..length-1]} in the text indexed by {\tt index}. It also
allocates {\tt *occ} (which must be freed by the caller) that
contains the locations of the {\tt *numocc} occurrences, in
arbitrary order.

\end{itemize}

\item {\bf Accessing the indexed text}
\begin{itemize}
\item {\tt int extract (void *index, ulong from, ulong to, uchar **snippet, ulong} \\
{\tt *snippet\_length)} allocates {\tt snippet} (which must be
freed by the caller) and writes the substring {\tt text[from..to]}
into it. Returns in {\tt snippet\_length} the length of the text
snippet actually extracted, since this could be less than {\tt
to-from+1} if {\tt to} is larger than the text size.

\item {\tt int display (void *index, uchar *pattern, ulong length, ulong numc, ulong *numocc,
uchar **snippet\_text, ulong **snippet\_lengths)} displays the
text snippet surrounding every occurrence of the substring {\tt
pattern[0..length-1]} within the indexed text. The snippet must
include {\tt numc} symbols before and after the pattern
occurrence, totalizing {\tt length+2*numc} symbols, or less if the
text boundaries are reached. The number of pattern occurrences is
stored in {\tt numocc}, and their snippets are stored in the
arrays {\tt snippet\_text} and {\tt snippet\_lengths} (which must
be freed by the caller). The first array is a symbol array of {\tt
numocc*(length+2*numc)} symbols, with a new snippet starting at
every multiple of {\tt length+2*numc}. The second array contains
integers, each indicating the real length of each of the {\tt
numocc} extracted snippets.

\item {\tt int length (void *index, ulong *length)} obtains the length of the
text indexed by index, in bytes.

\end{itemize}
\end{itemize}
}

\subsection{Texts}
\label{exp:texts}

We have chosen the texts forming the {\em Pizza\&Chili} collection
by following three basic considerations. First, we wished to cover
a representative set of application areas where the problem of
full-text indexing might be relevant, and for each of them we
selected texts freely available on the Web. Second, we aimed at
having one file per text type in order to avoid unreadable tables
of many results. Third, we have chosen the size of the texts to be
large enough in order to make indexing relevant and compression
apparent. These are the current collections provided in the
repository:

\begin{itemize}
\item \dna\ (DNA sequences). This file contains bare DNA sequences without 
descriptions, separated by {\tt newline}, obtained
from files available at the Gutenberg Project site: 
namely, from \textit{01hgp10} to
\textit{21hgp10}, plus \textit{0xhgp10} and \textit{0yhgp10}. Each
of the four DNA bases is coded as an uppercase letter A,G,C,T, and
there are a few occurrences of other special symbols.

\item \english\ (English texts). This file is the concatenation of English
texts selected from the collections
\textit{etext02}---\textit{etext05} available at the Gutenberg
Project sitei. 
We deleted the headers
related to the project so as to leave just the real text.

\item \pitches\ (MIDI pitch values). This file is a sequence of pitch
values (bytes whose values are in the range 0-127, plus a few
extra special values) obtained from a myriad of MIDI files freely
available on the Internet. The MIDI files were converted into the
IRP format by using the {\tt semex} tool by Kjell Lemstrom
\cite{semex-paper}. This is a human-readable tuple format, where
the 5th column is the pitch value. The pitch values were coded in
one byte each and concatenated all together.

\item \proteins\ (protein sequences). This file contains bare protein sequences
without descriptions, separated by {\tt newline}, obtained from the Swissprot 
database ({\small \tt ftp.ebi.ac.uk/ pub/databases/swissprot/}). Each of the 20
amino acids is coded as an uppercase letter.

\item \sources\ (source program code). This file is formed by C/Java
source codes obtained by concatenating all the \textit{.c},
\textit{.h}, \textit{.C} and \textit{.java} files of the
\textit{linux-2.6.11.6} ({\small \tt ftp.kernel.org}) and \textit{gcc-4.0.0}
({\small \tt ftp.gnu.org}) distributions.

\item  \xml\ (structured text). This file is in XML format and provides
bibliographic information on major computer science journals and
proceedings. It was downloaded from the DBLP archive at {\small \tt
dblp.uni-trier.de}.
\end{itemize}

For the experiments we have limited the short file \pitches\ to
its initial 50 MB, whereas all the other long files have been cut
down to their initial 200 MB. We show now some statistics on those
files. These statistics and the tools used to compute them are
available at the {\em Pizza\&Chili} site.

Table \ref{table:statistics} summarizes some general
characteristics of the selected files. The last column, inverse
match probability, is the reciprocal of the probability of
matching between two randomly chosen text symbols. This may be
considered as a measure of the {\em effective} alphabet size ---
indeed, on a uniformly distributed text, it would be precisely the
alphabet size.

\begin{table}[tb]
\caption{General statistics for our indexed texts.}
\label{table:statistics}
\centering
\begin{tabular}{|l|r|r|r|}
\hline
Text & Size (MB)& Alphabet size & Inv. match prob.\\
\hline \hline
\dna\     & $200$ & $16$  & $3.86$\\
\hline
\english\ & $200$ & $225$ & $15.12$\\
\hline
\pitches\ & $50$  & $133$ & $40.07$\\
\hline
\proteins\ & $200$ & $25$  & $16.90$\\
\hline \sources\ & $200$ & $230$ & $24.81$\\
\hline
\xml\    & $200$ & $96$  & $28.65$\\
\hline
\end{tabular}
\end{table}

Table \ref{table:entropies} provides some information about the
compressibility of the texts by reporting the value of $\hk{k}$
for $0 \le k \le 4$, measured as number of bits per input symbol.
As a comparison on the {\em real} compressibility of these texts,
Table \ref{table:compressors} shows the performance of three
well-known compressors (sources available in the site): \gzip\
(Lempel-Ziv-based compressor), \bzip\ (BWT-based compressor), and
\ppm\ ($k$-th order modeling compressor). Notice that, as $k$
grows, the value of $\hk{k}$ decreases but the size of the {\em
dictionary} of length-$k$ contexts grows significantly, eventually
approaching the size of the text to be compressed. Typical values
of $k$ for \ppm\ are around 5 or 6. It is interesting to note in
Table \ref{table:compressors} that the compression ratios
achievable by the tested compressors may be superior to $\hk{4}$,
because they use (explicitly or implicitly) longer contexts.

\begin{table}[tb]
\caption{Ideal compressibility of our indexed texts. For every
$k$-th order model, with $0 \le k \le 4$, we report the number of
distinct contexts of length $k$, and the empirical entropy
$\hk{k}$, measured as number of bits per input symbol.}
\label{table:entropies}
\centering
\begin{scriptsize}
\begin{tabular}{|l|c|c|c|r|c|r|c|r|c|r|}
\hline
& & & \multicolumn{2}{|c|}{1st order} & \multicolumn{2}{|c|}{2nd order} & \multicolumn{2}{|c|}{3rd order} & \multicolumn{2}{|c|}{4th order}\\
\cline{4-11}
Text & $\log \sigma$  & $\hk{0}$ & $\hk{1}$ & \#~ & $\hk{2}$ & \#~~~ & $\hk{3}$ &
\#~~~~ & $\hk{4}$ & \#~~~~~~\\
\hline \hline
\dna     & $4.000$ & $1.974$ & $1.930$ & $16$ & $1.920$ & $152$ & $1.916$ & $683$ & $1.910$ & $2222$\\
\hline
\english & $7.814$ & $4.525$ & $3.620$ & $225$ & $2.948$ & $10829$ & $2.422$ & $102666$ & $2.063$ & $589230$\\
\hline
\pitches & $7.055$ & $5.633$ & $4.734$ & $133$ & $4.139$ & $10946$ & $3.457$ & $345078$ & $2.334$ & $3845792$\\
\hline
\proteins & $4.644$ & $4.201$ & $4.178$ & $25$ & $4.156$ & $607$ & $4.066$ & $11607$ & $3.826$ & $224132$\\
\hline
\sources & $7.845$ & $5.465$ & $4.077$ & $230$ & $3.102$ & $9525$ & $2.337$ & $253831$ & $1.852$ & $1719387$\\
\hline
\xml    & $6.585$ & $5.257$ & $3.480$ & $96$ & $2.170$ & $7049$ & $1.434$ & $141736$ & $1.045$ & $907678$\\
\hline
\end{tabular}
\end{scriptsize}
\end{table}

\begin{table}[tb]
\caption{Real compressibility of our indexed texts, as achieved by
the best-known compressors: \gzip\ (option {\tt -9}), \bzip\
(option {\tt -9}), and \ppm\ (option {\tt -l 9}).}
\label{table:compressors}
\centering
\begin{tabular}{|l|c|c|c|c|}
\hline
Text & $\hk{4}$ & \gzip\ &\bzip\ & \ppm \\
\hline \hline
\dna     & $1.910$ & $2.162$ & $2.076$ & $1.943$\\
\hline
\english & $2.063$ & $3.011$ & $2.246$ & $1.957$\\
\hline
\pitches & $2.334$ & $2.448$ & $2.890$ & $2.439$\\
\hline
\proteins & $3.826$ & $3.721$ & $3.584$ & $3.276$\\
\hline
\sources & $1.852$ & $1.790$ & $1.493$ & $1.016$\\
\hline
\xml    & $1.045$ & $1.369$ & $0.908$ & $0.745$\\
\hline
\end{tabular}
\end{table}

\section{Experimental Results}
\label{sec:exper}

In this section we report experimental results from a 
subset of the compressed indexes available at the {\em
Pizza\&Chili} site. All the experiments were executed on a $2.6$
GHz Pentium 4, with $1.5$ GB of main memory, and running Fedora
Linux. The searching and building algorithms for all compressed
indexes were coded in C/C++ and compiled with {\tt gcc} or {\tt
g++} version $4.0.2$.

We restricted our experiments to a few indexes: Succinct Suffix
Array (version \ssa{\tt \_v2} in {\em Pizza\&Chili}),
Alphabet-Friendly FM-index (version \afindex{\tt \_v2}
in {\em Pizza\&Chili}), Compressed Suffix Array (\csa\ in {\em Pizza\&Chili}),
and LZ-index (version \lzindex{\tt 4} in {\em
Pizza\&Chili}), because they are the best representatives of the
three classes of compressed indexes we discussed in
Section~\ref{sec:indexes}. This small number will provide us with
a succinct, yet significant, picture of the performance of all
known compressed indexes \cite{NMacmcs06}.

There is no need to say that further algorithmic engineering of
the indexes experimented in this paper, as well of the other
indexes available in the {\em Pizza\&Chili} site, could possibly
change the charts and tables shown below. However, we believe that
the overall conclusions drawn from our experiments should not
change significantly, unless new algorithmic ideas are devised for
them. Indeed, the following list of experimental results has a
twofold goal: on one hand, to quantify the space and time
performance of compressed indexes over real datasets, and on the
other hand, to motivate further algorithmic research by
highlighting the limitations of the present indexes and their
implementations.

\subsection{Construction}\label{subsec:construction}

Table~\ref{table:parameters} shows the parameters used to
construct the indexes in our experiments.
Table~\ref{table:construction} shows construction time and space
for one collection, namely \english, as all the others give roughly
similar results. The bulk of the time of \ssa\ and \csa\ is
that of suffix array construction (prior to its compression). The
times differ because different suffix array construction
algorithms are used (see Section \ref{sub:indexes}). The \afindex\
takes much more time because it needs to run the compression
boosting algorithm over the suffix array. The \lzindex\ spends
most of the time in parsing the text and creating the LZ78 and
reverse tries. In all cases construction times are practical, 1--4
sec/MB with our machine.

\begin{table}[tb]
\caption{Parameters used for the different indexes in our experiments. The
cases of multiple values correspond to space/time tradeoff curves.}
\label{table:parameters}
\centering
\begin{tabular}{|l|c|c|}
\hline
Index & count & locate / extract \\
\hline \hline
\afindex & $-$ & $s_A=\{4,16,32,64,128,256\}$ \\
\hline
\csa     & $s_\psi=\{128\}$ & $s_A=\{4,16,32,64,128,256\};s_\psi=\{128\}$ \\
\hline
\lzindex & $\epsilon = \{\frac{1}{4}\}$ & $\epsilon =\{1, \frac{1}{2}, \frac{1}{3}, \frac{1}{4}, \frac{1}{5}, \frac{1}{10}, \frac{1}{20}\}$\\
\hline
\ssa     &$-$ & $s_A=\{4,16,32,64,128,256\}$ \\
\hline
\end{tabular}
\end{table}

The memory usage might be problematic, as it is 5--9 times the
text size. Albeit the final index is small, one
needs much memory to build it first\footnote{In particular, this limited
us to indexing up to 200 MB of text in our machine.}. This is a problem of
compressed indexes, which is attracting a lot of practical and
theoretical research
\cite{LSSY02_Suffix,ANisaac05,HSS03_breaking,MN06}.

\begin{table}[tb]
\caption{Time and peak of main memory usage required to build the
various indexes over the 200 MB file \english. The indexes are
built using the default value for the locate tradeoff (that is,
$s_A=64$ for \afindex\ and \ssa; $s_A=64$ and $s_\psi=128$ for
\csa; and $\epsilon = \frac{1}{4}$ for the \lzindex).}
\label{table:construction}
\centering
\begin{tabular}{|l|c|c|}
\hline
Index & Build Time (sec) & Main Memory Usage (MB)\\
\hline \hline
\afindex    & $772$ & $1,751$\\
\hline
\csa        & $423$ & $1,801$\\
\hline
\lzindex    & $198$ & $1,037$\\
\hline
\ssa        & $217$ & $1,251$\\
\hline
\end{tabular}

\end{table}

We remark that the indexes allow different space/time tradeoffs.
The \ssa\ and \afindex\ have a sampling rate parameter $s_A$ that
trades locating and extracting time for space. More precisely,
they need $O(s_A)$ accesses to the wavelet tree for locating, and
$O(s_A +r-l+1)$ accesses to extract $T_{l,r}$, in exchange for
$\frac{n\log n}{s_A}$ additional bits of space. We can remove
those structures if we are only interested in counting.

The \csa\ has two space/time tradeoffs. A first one, $s_\psi$,
governs the access time to $\psi$, which is $O(s_\psi)$ in
exchange for $\frac{n\log n}{s_\psi}$ bits of space required by
the samples. The second, $s_A$, affects locating and extracting
time just as above. For pure counting we can remove the sampling
related to $s_A$, whereas for locating the best is to use the
default value (given by Sadakane) of $s_\psi=128$. The best choice
for extracting is less clear, as it depends on the length of
the substring to extract.

Finally, the \lzindex\ has one parameter $\epsilon$ which trades
counting/locating time per space occupancy: The cost per
candidate occurrence is multiplied by $\frac{1}{\epsilon}$, and
the additional space is $2\epsilon n H_k(T)$ bits. No structure
can be removed in the case of counting, but space can be halved if
the extract operation is the only one needed (just remove the
reverse trie).

\subsection{Counting}

We searched for $50,000$ patterns of length $m=20$, randomly
chosen from the indexed texts. The average counting time was then
divided by $m$ to display counting time per symbol. This is
appropriate because the counting time of the indexes is linear in
$m$, and 20 is sufficiently large to blur small constant
overheads. The exception is the \lzindex, whose counting time is 
superlinear in $m$, and not competitive at all for this task.

\begin{table}
\caption{Experiments on the counting of pattern occurrences. Time
is measured in microseconds per pattern symbol. The space usage is
expressed as a fraction of the original text size. We put in boldface those 
results that lie within 10\% of the best space/time tradeoffs.}
\label{table:counttime}
\centering
\begin{scriptsize}
\begin{tabular}{|l|r|r|r|r|r|r|r|r|r|r|}
\hline
&\multicolumn{2}{|c}{\ssa} &\multicolumn{2}{|c|}{\afindex} &
\multicolumn{2}{c|}{\csa} & \multicolumn{2}{c|}{\lzindex}& \multicolumn{2}{c|}{\sau}\\
\cline{2-11}
\hspace{-3pt}Text     \hspace{-3pt}   & Time & Space & Time & Space & Time & Space & Time & Space  & Time & Space\\
\hline
\hline
\hspace{-3pt}\dna     \hspace{-3pt}   & \bf 0.956 &  \bf 0.29 &   $1.914$  & \bf 0.28  &   $5.220$  &   $0.46$  &   $43.896$  & $0.93$ & $0.542$ & $5$ \\ \hline
\hspace{-3pt}\english \hspace{-3pt}   & \bf 2.147 &   $0.60$ &   $2.694$  & \bf 0.42  &   $4.758$  & \bf 0.44  &   $68.774$  & $1.27$ & $0.512$ & $5$ \\ \hline
\hspace{-3pt}\pitches \hspace{-3pt}   & \bf 2.195 &   $0.74$ &   $2.921$  & \bf 0.66  &   $3.423$  & \bf 0.63  &   $55.314$  & $1.95$ & $0.363$ & $5$ \\ \hline
\hspace{-3pt}\proteins\hspace{-3pt}   & \bf 1.905 & \bf 0.56 &   $3.082$  & \bf 0.56  &   $6.477$  &   $0.67$  &   $47.030$  & $1.81$ & $0.479$ & $5$ \\ \hline
\hspace{-3pt}\sources \hspace{-3pt}   & \bf 2.635 &   $0.72$ &   $2.946$  &   $0.49$  &   $4.345$  & \bf 0.38  &   $162.444$ & $1.27$ & $0.499$ & $5$ \\ \hline
\hspace{-3pt}\xml     \hspace{-3pt}   & $2.764$   &   $0.69$ & \bf 2.256  &   $0.34$  &   $4.321$  & \bf 0.29  &   $306.711$ & $0.71$ & $0.605$ & $5$ \\ \hline
\end{tabular}
\end{scriptsize}
\end{table}

Table \ref{table:counttime} shows the results on this test. The
space of the \ssa, \afindex, and \csa\ does not include what is
necessary for locating and extracting. We can see that, as
expected, the \afindex\ is always smaller than the \ssa, yet they
are rather close on \dna\ and \proteins\ (where the zero-order
entropy is not much larger than higher-order entropies). The space
usages of the \afindex\ and the \csa\ are similar and usually the best, 
albeit the \csa\ predictably loses in counting time on smaller alphabets 
(\dna, \proteins), due to its $O(m\log n)$ rather than $O(m\log\sigma)$ 
complexity. The \csa\ takes advantage of larger alphabets with
good high-order entropies (\sources, \xml), a combination where
the \afindex, despite of its name, profits less. Note that the
space performance of the \csa\ on those texts confirms that its
space occupancy may be below the zero-order entropy.

With respect to time, the \ssa\ is usually the fastest thanks to
its simplicity. Sometimes the \afindex\ gets close and it is
actually faster on \xml. The \csa\ is rarely competitive
for counting, and the \lzindex\ is well out of bounds for
this experiment. Notice that the plain suffix array (last column
in Table \ref{table:counttime}) is 2--6 times faster than any
compressed index, but its space occupancy can be up to 18 times
larger.


\subsection{Locate}

\begin{figure}[p]
\centering
\begin{tabular}{c}
\begin{tabular}{cc}
\includegraphics[width=0.45\textwidth]{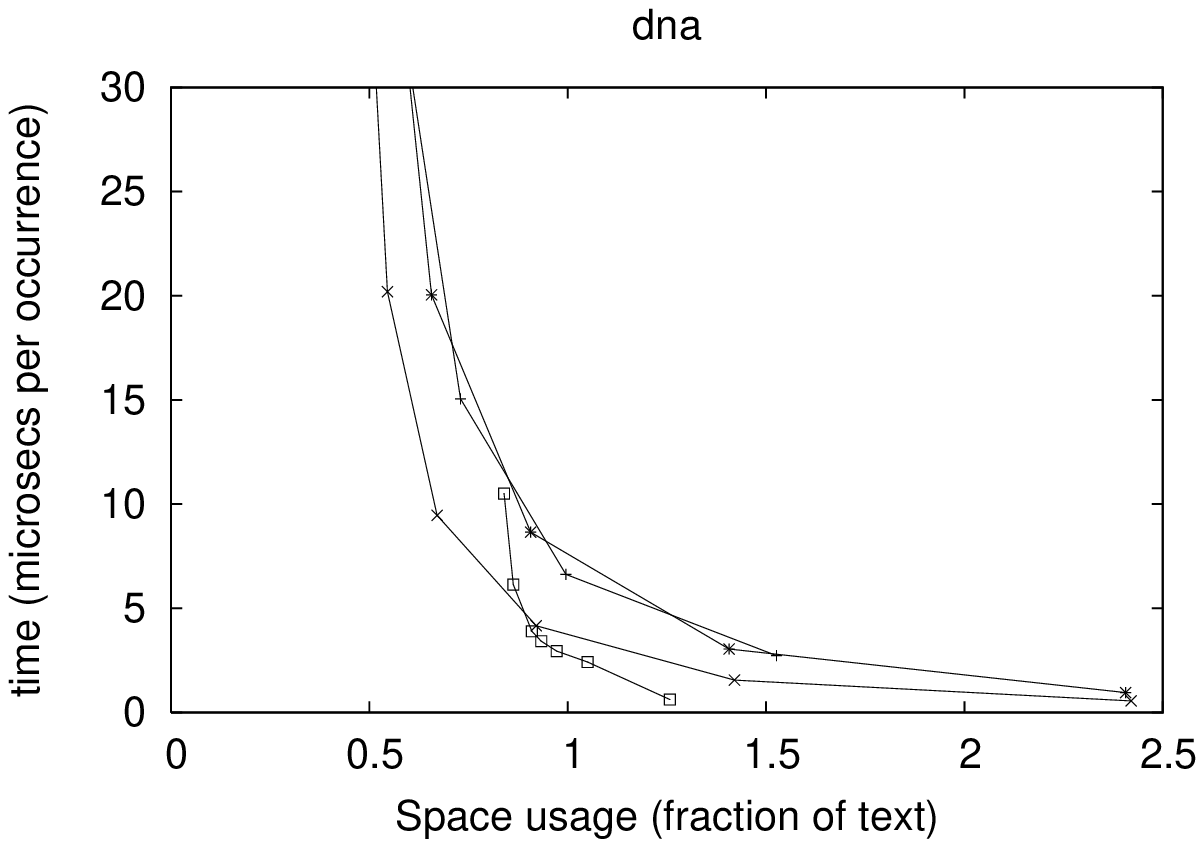} &
\includegraphics[width=0.45\textwidth]{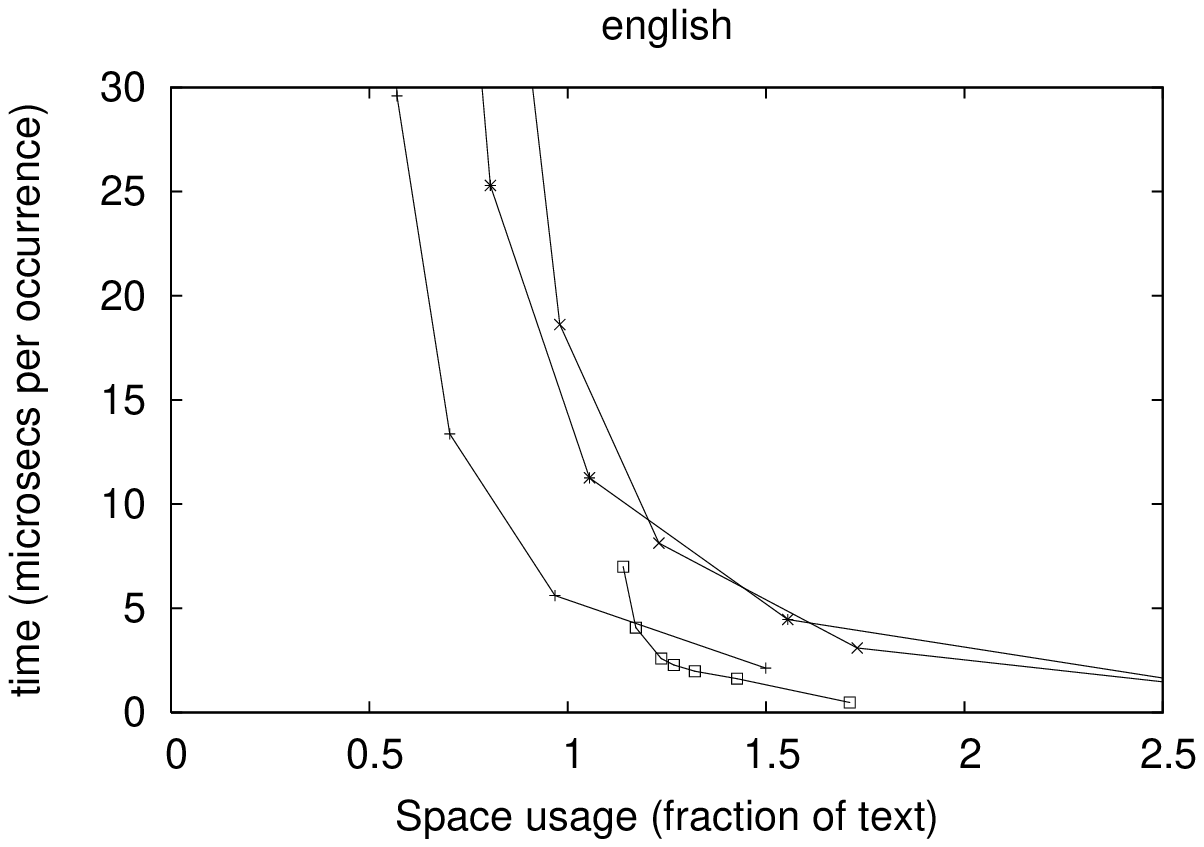} \\ 
\includegraphics[width=0.45\textwidth]{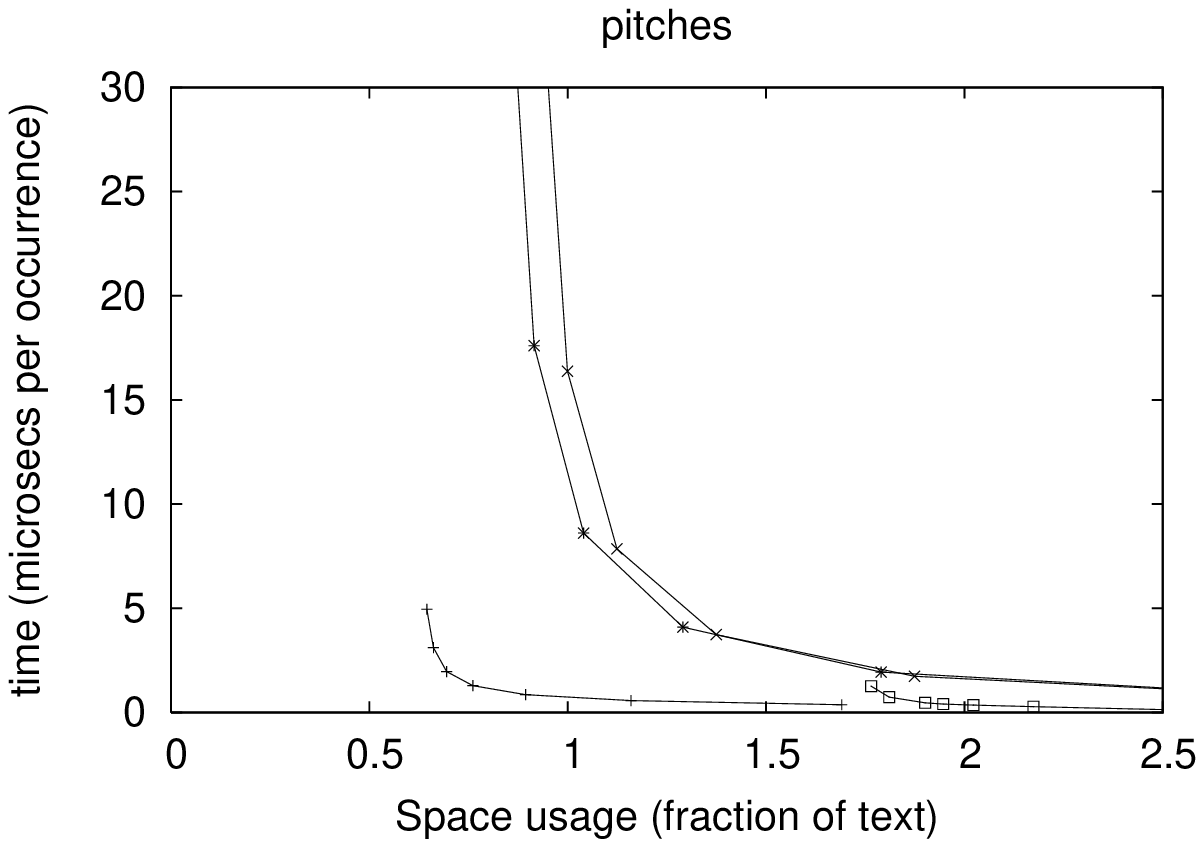} &
\includegraphics[width=0.45\textwidth]{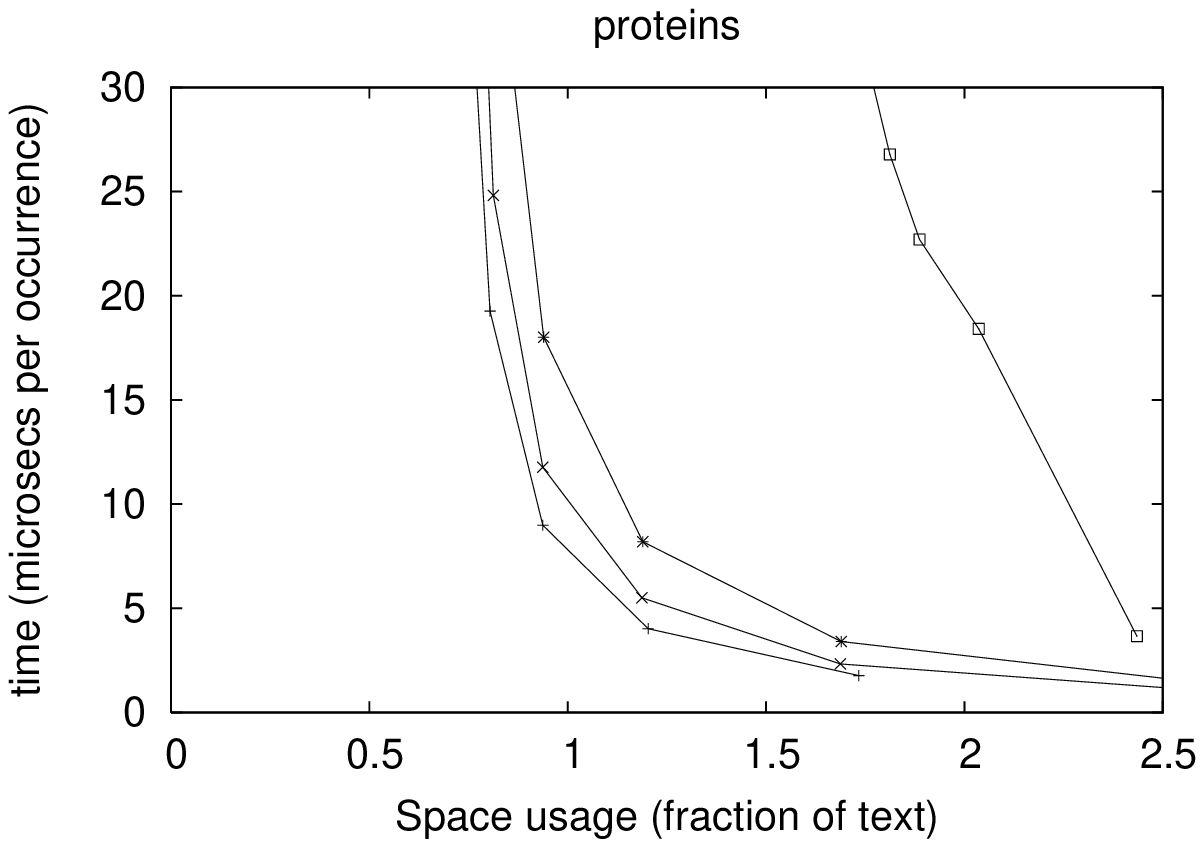} \\ 
\includegraphics[width=0.45\textwidth]{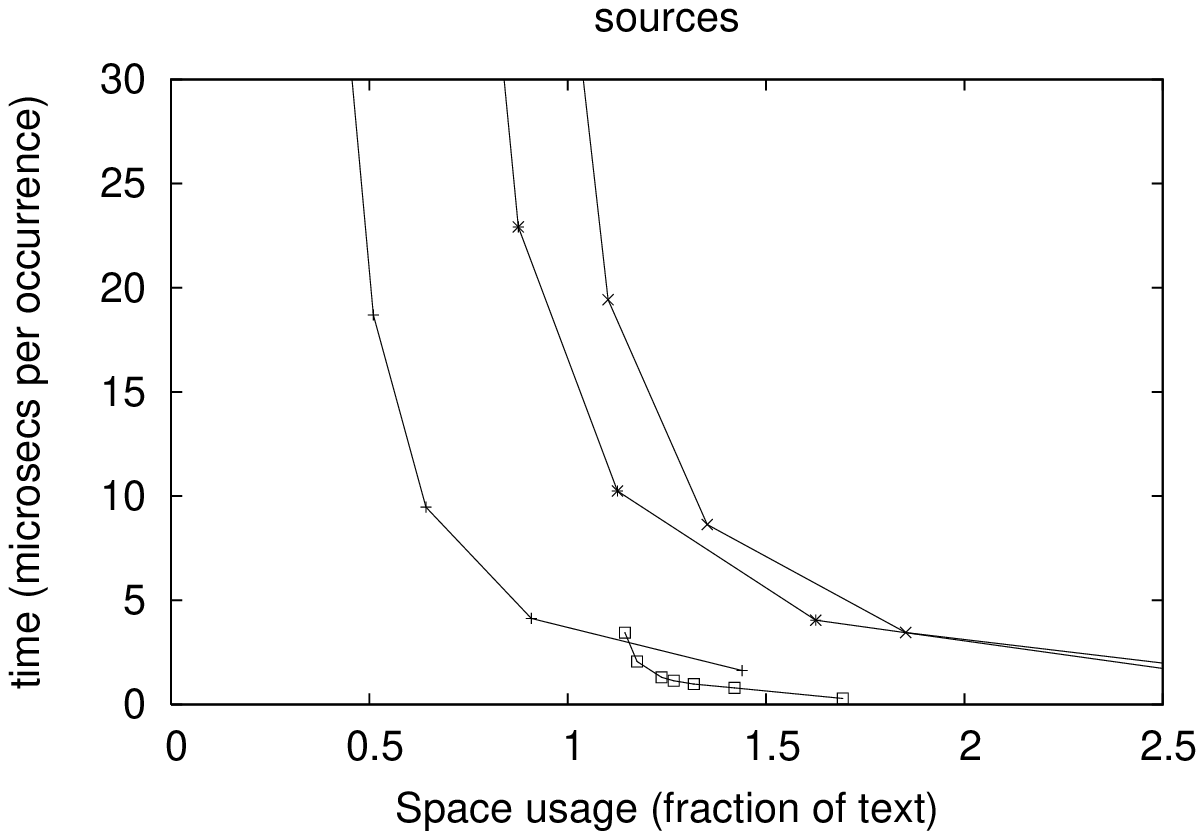} &
\includegraphics[width=0.45\textwidth]{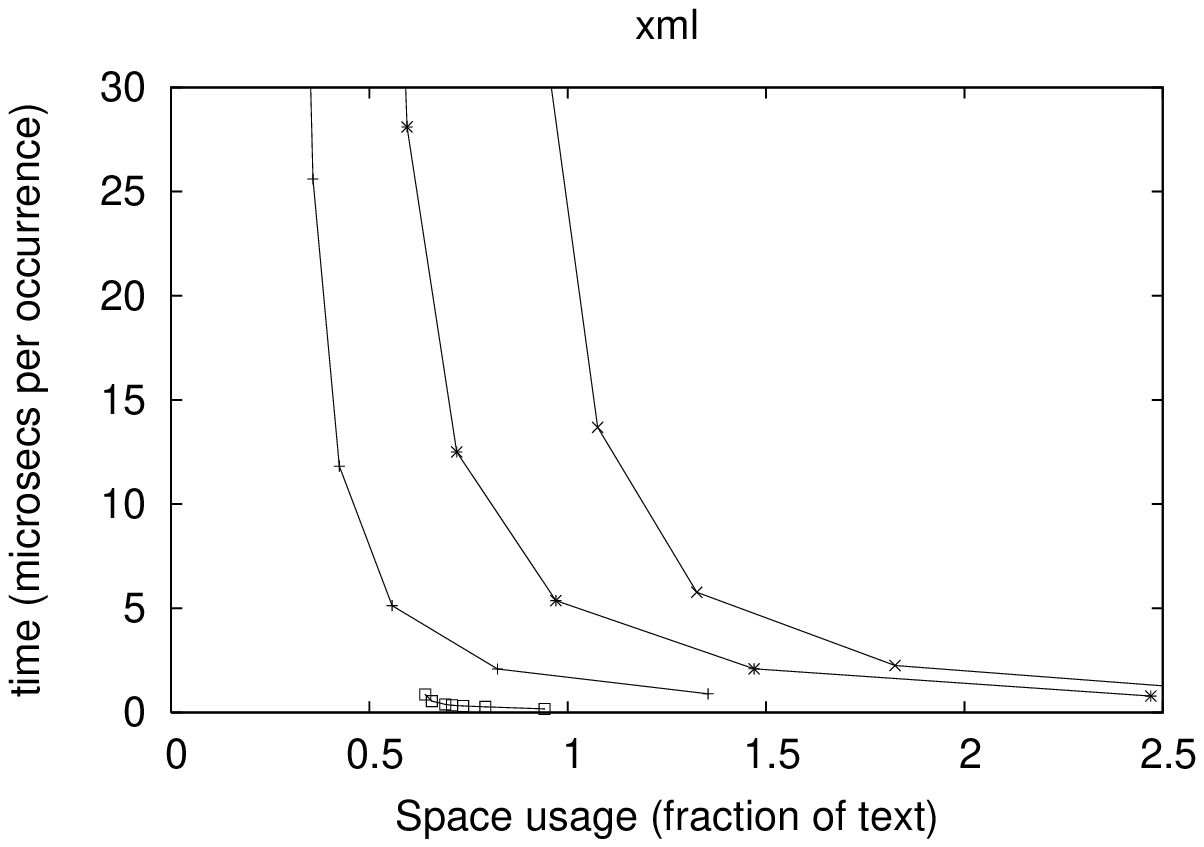} \\
\end{tabular}\\\\
\begin{tabular}{|lcclc|}
\hline
{{\tiny $*$}} &  {\afindex} && {{\tiny $+$}} & {\csa} \\
{{\tiny $\Box$}} & {\lzindex} && {{\tiny $\times$}} & {\ssa} \\
\hline
\end{tabular}
\end{tabular}
\caption{Space-time tradeoffs for locating occurrences of patterns of length 5.}
\label{fig:plotlocate}
\end{figure}

We locate sufficient random patterns of length $5$ to obtain a
total of 2--3 million occurrences per text (see
Table~\ref{table:locpatterns}). This way we are able to evaluate
the average cost of a single locate operation, by making the
impact of the counting cost negligible.
Fig. \ref{fig:plotlocate} reports the time/space tradeoffs
achieved by the different indexes for the locate operation.

\begin{table}
\caption{Number of searched patterns of length 5 and total number of located
occurrences.}
\label{table:locpatterns}
\centering
\begin{tabular}{|l|r|r|}
\hline
Text & \# patterns & \# occurrences\\
\hline \hline
\dna     & $10$  & $2,491,410$\\
\hline
\english & $100$& $2,969,876$\\
\hline
\pitches & $200$& $2,117,347$\\
\hline
\proteins& $3,500$ & $2,259,125$\\
\hline
\sources & $50$ & $2,130,626$\\
\hline
\xml    & $20$ & $2,831,462$\\
\hline
\end{tabular}
\end{table}

We remark that the implemented indexes include the sampling
mechanism for locate and extract as a single module, and therefore
the space for both operations is included in these plots.
Therefore, the space could be reduced if we only wished to locate.
However, as extracting snippets of pattern occurrences is an
essential functionality of a self-index, we consider that the
space for efficient extraction should always be
included.\footnote{Of course, we could have a sparser sampling for
extraction, but we did not want to complicate the evaluation more
than necessary.}

The comparison shows that usually \csa\ can achieve the best
results with minimum space, except on \dna\ where the \ssa\
performs better as expected (given its query time complexity, see
before), and on \proteins\ for which the suffix-array-based indexes perform 
similarly (and the \lzindex\ does much worse). The 
\csa\ is also the most attractive alternative if we fix that the
space of the index should be equal to that of the text (recall
that it includes the text), being the exceptions \dna\ and \xml,
where the \lzindex\ is superior.

The \lzindex\ can be much faster than the others if one is willing
to pay for some extra space. The exceptions are \pitches, where
the \csa\ is superior, and \proteins, where the \lzindex\ performs
poorly. This may be caused by the large number of patterns that
were searched to collect the 2--3 million occurrences (see
Table~\ref{table:locpatterns}), as the counting is expensive on
the \lzindex.

\begin{table}[tb]
\caption{Locate time required by \sau\ in microseconds per
occurrence, with $m=5$. We recall that this implementation requires $5$ bytes
per indexed symbol.} 
\label{table:classiclocate}
\centering
\begin{tabular}{|l|r|r|r|r|r|r|}
\hline
& \dna & \english & \pitches & \proteins & \sources & \xml \\
\hline \hline
\sau & 0.005 & 0.005 & 0.006 & 0.007 & 0.007 & 0.006 \\
\hline
\end{tabular}
\end{table}

Table \ref{table:classiclocate} shows the locate time required by
an implementation of the classical suffix array: it is between 100
and 1000 times faster than any compressed index, but always 5
times larger than the indexed text. Unlike counting, where
compressed indexes are comparable in time with classical ones,
locating is much slower on compressed indexes. This comes from the
fact that each locate operation (except on the \lzindex) requires
to perform several random memory accesses, depending on the
sampling step. In contrast, all the occurrences are contiguous in
a classical suffix array. As a result, the compressed indexes are
currently very efficient in case of selective queries, but
traditional indexes become more effective when locating many
occurrences. This fact has triggered recent research activity on
this subject (e.g., \cite{GNcpm07.1}) but a deeper understanding
on index performance on hierarchical memories is still needed.


\subsection{Extract}

\begin{figure}[p]
\centering
\begin{tabular}{c}
\begin{tabular}{cc}
\includegraphics[width=0.45\textwidth]{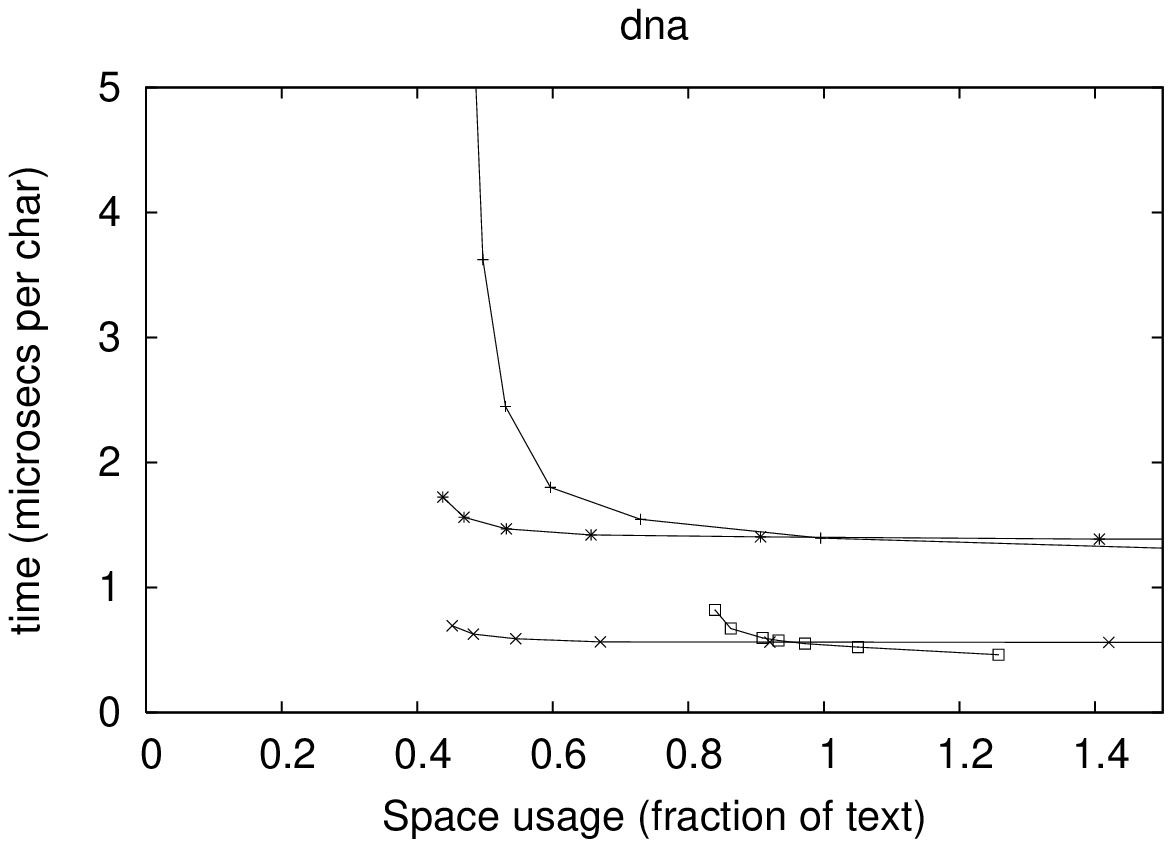} &
\includegraphics[width=0.45\textwidth]{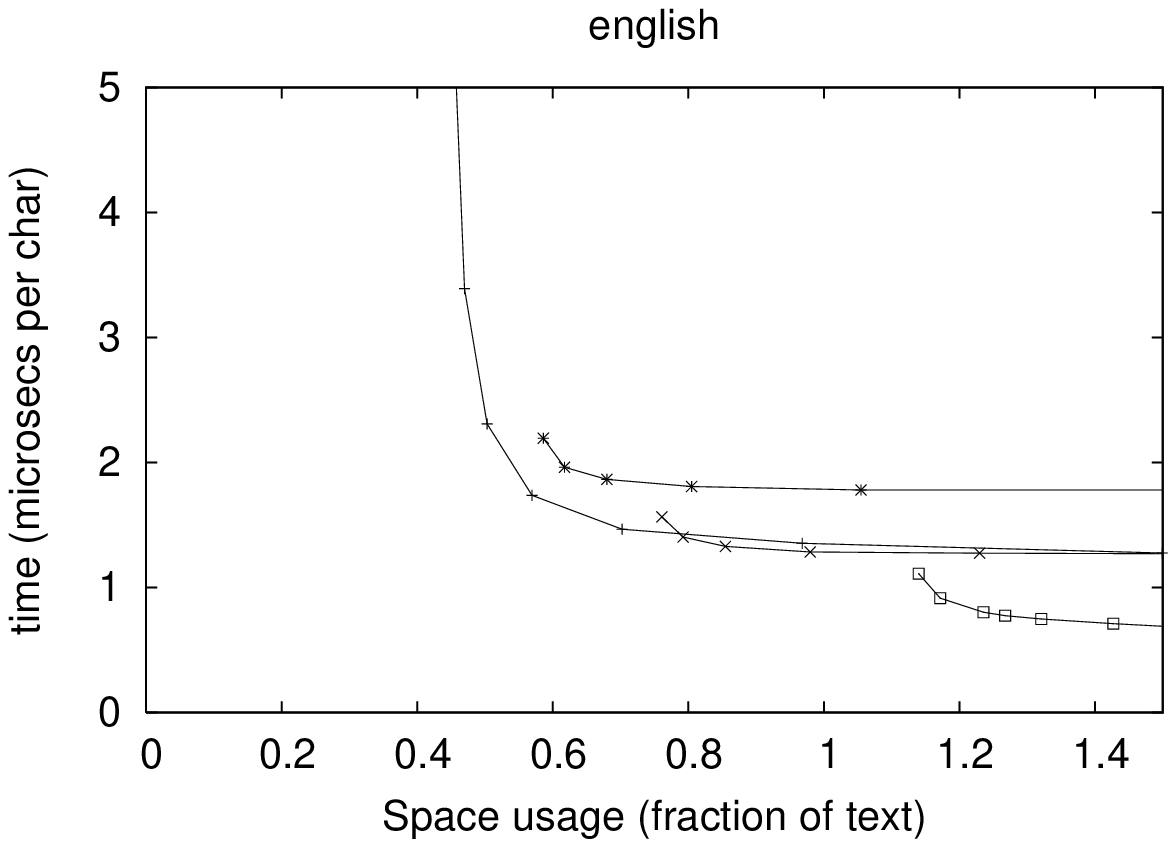} \\ 
\includegraphics[width=0.45\textwidth]{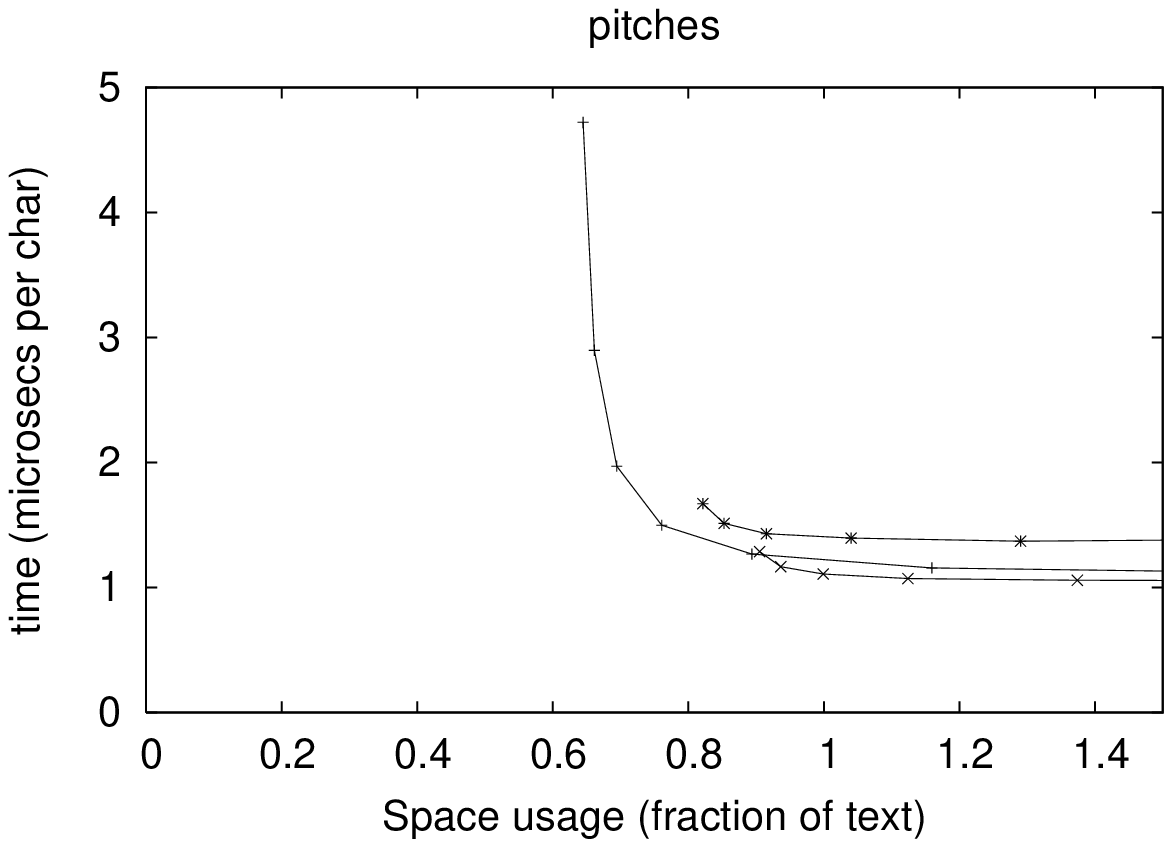} &
\includegraphics[width=0.45\textwidth]{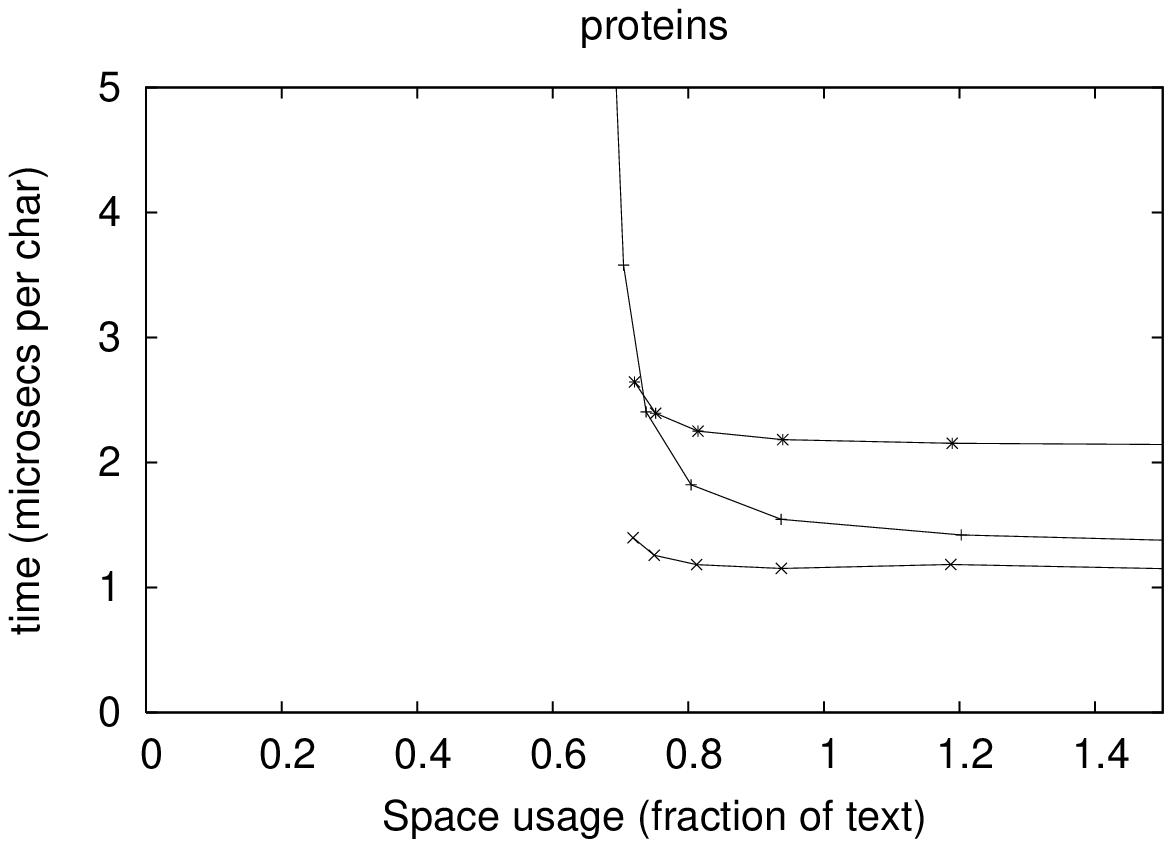} \\ 
\includegraphics[width=0.45\textwidth]{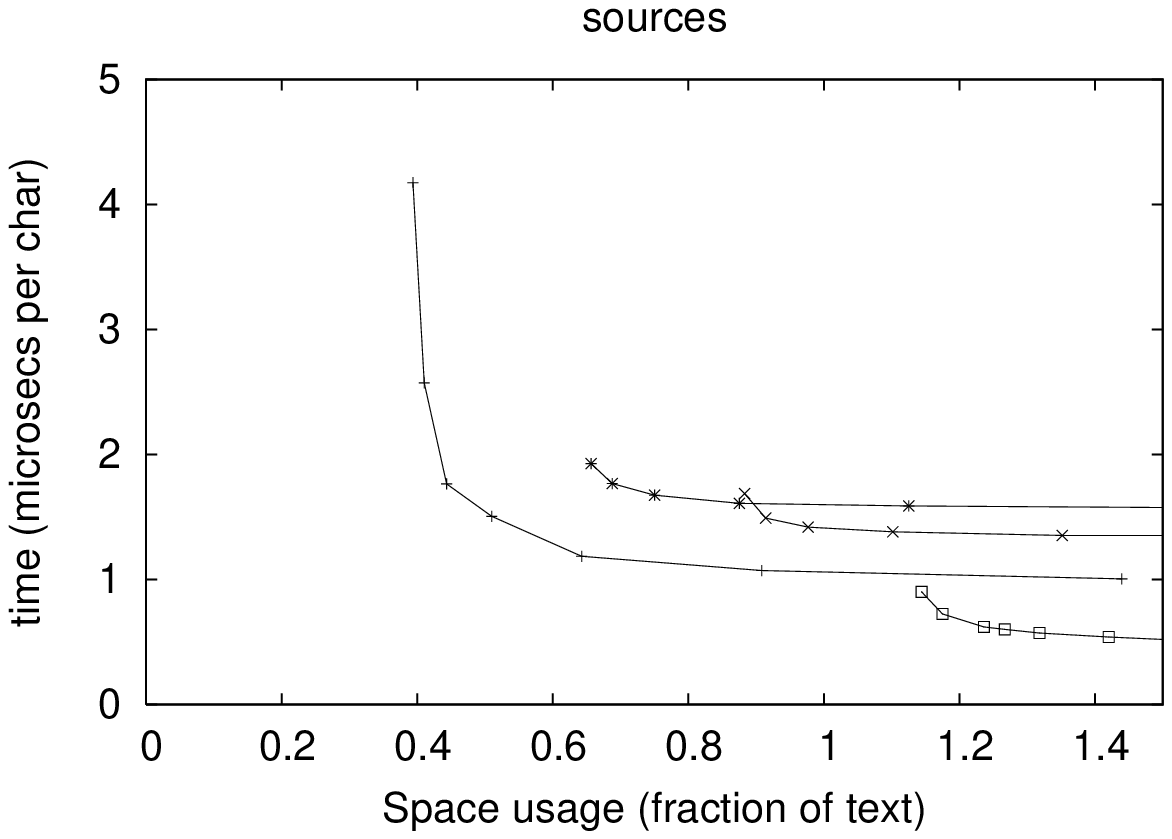} &
\includegraphics[width=0.45\textwidth]{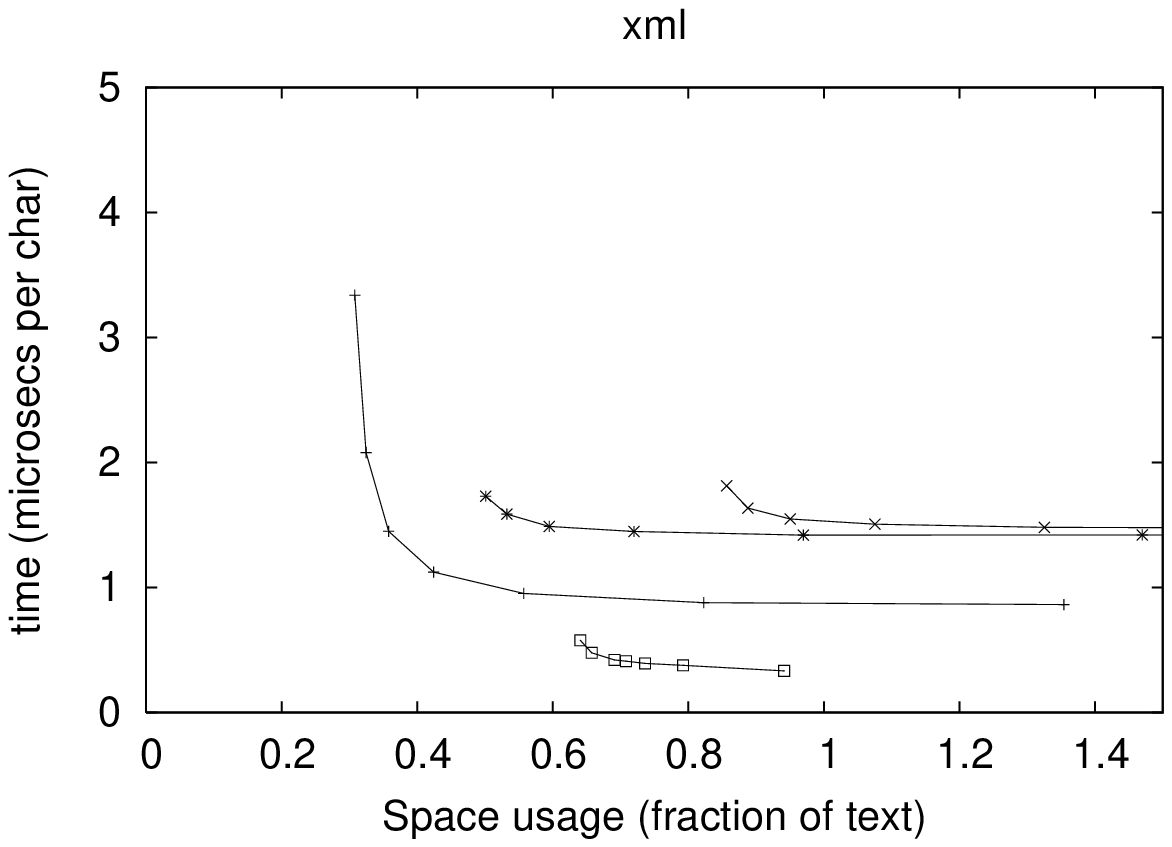} \\
\end{tabular}\\\\
\begin{tabular}{|lcclc|}
\hline
{{\tiny $*$}} &  {\afindex} && {{\tiny
$+$}} & {\csa} \\
{{\tiny $\Box$}} & { \lzindex} && {{\tiny
$\times$}} & {\ssa} \\
\hline
\end{tabular}
\end{tabular}
\caption{Space-time tradeoffs for extracting text symbols.}
\label{fig:plotextract}
\end{figure}

We extracted substrings of length $512$ from random text
positions, for a total of 5 MB of extracted text.
Fig. \ref{fig:plotextract} reports the time/space tradeoffs
achieved by the tested indexes. We still include both space to
locate and extract, but we note that the sampling step affects
only the time to reach the text segment to extract from the
closest sample, and afterwards the time is independent of the
sampling. We chose length 512 to smooth out the effect of this
sampling.

The comparison shows that, for extraction purposes, the \csa\ is
better for \sources\ and \xml, whereas the \ssa\ is better on
\dna\ and \proteins. On \english\ and \pitches\ both are rather
similar, albeit the \csa\ is able to operate on reduced space. On
the other hand, the \lzindex\ is much faster than the others on
\xml, \english\ and \sources, if one is willing to pay some
additional space.\footnote{Actually the \lzindex\ is not plotted
for \pitches\ and \proteins\ because it needs more than 1.5 times
the text size.}

It is difficult to compare these times with those of a classical
index, because the latter has the text readily available.
Nevertheless, we note that the times are rather good: using the
same space as the text (and some times up to half the space) for
all the functionalities implemented, the compressed indexes are
able to extract around 1 MB/sec, from arbitrary positions. This
shows that self-indexes are appealing as compressed-storage
schemes with the support of random accesses for snippet
extraction.

\section{Conclusion and Future Work}
\label{sec:concl}

In this paper we have addressed the new fascinating technology of
compressed text indexing. We have explained the main principles
used by those indexes in practice, and presented the {\em
Pizza\&Chili} site, where implementations and testbeds are readily
available for use. Finally, we have presented experiments that
demonstrate the practical relevance of this emerging technology.
Table~\ref{table:summary} summarizes our experimental results by
showing the most promising compressed index(es) depending on the
text type and task.

\begin{table}[ht]
\caption{The most promising indexes given the size and time they
obtain for each operation/text.} 
\label{table:summary}
\centering
\begin{tabular}{|l|r|r|r|r|r|r|}
\hline
& \dna & \english\ & \pitches\ & \proteins\ & \sources\ & \xml \\
\hline \hline
\multirow{2}{*}{count} & \ssa & \ssa & \afindex  & \ssa & \csa & \afindex \\
& - &  \afindex & \ssa & - &  \afindex & - \\ \hline
\multirow{2}{*}{locate} & \lzindex & \csa & \csa & \ssa & \csa & \csa \\
& \ssa & \lzindex & - & - & \lzindex & \lzindex \\ \hline
\multirow{2}{*}{extract} & \ssa & \csa & \csa & \ssa & \csa & \csa \\
& - & \lzindex & - & - & \lzindex & \lzindex \\
\hline
\end{tabular}
\end{table}

For counting the best indexes are \ssa\ and \afindex. This stems
from the fact that they achieve very good zero- or high-order
compression of the indexed text, while their average counting
complexity is $O(m H_0(T))$. The \ssa\ has the advantage of a
simpler search mechanism, but the \afindex\ is superior for texts
with small high-order entropy (i.e. \xml, \sources, \english). The
\csa\ usually loses because of its $O(m\log n)$ counting
complexity.

For locating and extracting, which are LF-computation intensive,
the \afindex\ is hardly better than the simpler \ssa\ because the
benefit of a denser sampling does not compensate for the presence of
many wavelet trees. The \ssa\ wins for small-alphabet data, like
\dna\ and \proteins. Conversely, for all other high-order
compressible texts the \csa\ takes over the other approaches. We
also notice that the \lzindex\ is a very competitive choice when extra
space is allowed and the texts are highly compressible.

The ultimate moral is that there is not a clear winner for all
text collections. Nonetheless, our results provide an {\em upper
bound} on what these compressed indexes can achieve in practice:

\begin{description}

\item[Counting.] We can compress the text within 30\%--50\% of its original size,
and search for 20,000--50,000 patterns of 20 chars each
within a second.

\item[Locate.] We can compress the text within 40\%--80\% of its original size,
and locate about 100,000 pattern occurrences per second.

\item[Extract.] We can compress the text within 40\%--80\% of its original size,
and decompress its symbols at a rate of about 1 MB/second.

\end{description}

The above figures are from one (count) to three (locate) orders of
magnitudes slower than what one can achieve with a plain suffix
array, at the benefit of using up to 18 times less space. This
slowdown is due to the fact that search operations in compressed
indexes access the memory in a non-local way thus eliciting many
cache/IO misses, with a consequent degradation of the overall time
performance. Nonetheless compressed indexes achieve a
(search/extract) throughput which is significant and may match the
efficiency specifications of most software tools which run on a
commodity PC. We therefore hope that this paper will spread their
use in any software that needs to process, store and mine text
collections of any size. Why using much space when squeezing and
searching is nowadays simultaneously affordable?

\bibliographystyle{plain}
\bibliography{survey}

\end{document}